\documentclass{aastex631}

\usepackage{gensymb}
\usepackage{hyperref}
\usepackage{wrapfig}

\definecolor{Gray}{gray}{0.9}
\shorttitle{Degree of Delta}
\shortauthors{Norton et al.}
\DeclareUnicodeCharacter{2212}{-}

\graphicspath{{./}{figures/}}

\begin{document}

\title{Characterizing the Umbral Magnetic Knots of $\delta$-Sunspots}

\author[0000-0003-2622-7310]{Aimee A. Norton}
\affiliation{HEPL Solar Physics, Stanford University, Stanford, CA 94305-4085}
\author[0000-0001-9640-6889]{Peter J. Levens}
\affiliation{HEPL Solar Physics, Stanford University, Stanford, CA 94305-4085}
\author[0000-0002-2544-2927]{Kalman J. Knizhnik}
\affiliation{Naval Research Laboratory, 4555 Overlook Ave SW, Washington, DC 20375}
\author[0000-0002-4459-7510]{Mark G. Linton}
\affiliation{Naval Research Laboratory, 4555 Overlook Ave SW, Washington, DC 20375}
\author[0000-0002-0671-689X]{Yang Liu}
\affiliation{HEPL Solar Physics, Stanford University, Stanford, CA 94305-4085}

\begin{abstract}
Delta ($\delta$)-spots are active regions (ARs) in which positive and negative umbrae share a penumbra. They are known to be the source of strong flares. We introduce a new quantity, the degree of $\delta$ (Do$\delta$), to measure the fraction of umbral flux participating in the $\delta$-configuration and to isolate the dynamics of the magnetic knot, i.e. adjacent umbrae in the $\delta$-configuration. Using Helioseismic and Magnetic Imager data, we analyze 19 $\delta$-spots and 11 $\beta$-spots in detail, and 120 $\delta$-spots in less detail.  We find that $\delta$-regions are not in a $\delta$-configuration for the entire time but spend 55$\%$ of their observed time as $\delta$-spots with an average, maximum Do$\delta$ of 72$\%$. Compared to $\beta$-spots,  $\delta$-spots have 2.6$\times$ the maximum umbral flux, 1.9$\times$ the flux emergence rate, 2.6$\times$ the rotation, and 72$\times$ the flare energy. On average, the magnetic knots rotate 17$\degree$ day$^{-1}$ while the $\beta$-spots rotate 2$\degree$ day$^{-1}$. Approximately 72$\%$ of the magnetic knots present anti-Hale or anti-Joy tilts, contrasting starkly with only 9$\%$ of the $\beta$-spots. A positive correlation exists between $\phi_{Do\delta}$ and the flare energy emitted by that region. The $\delta$-spots obey the hemispheric current helicity rule 64$\%$ of the time. 84$\%$ of the $\delta$-spots are formed by single flux emergence events and 58$\%$ have a quadrupolar magnetic configuration. The $\delta$-spot characteristics are consistent with the formation mechanism signatures as follows:  42$\%$ with the kink instability or Sigma effect, 32$\%$ with multi-segment buoyancy, 16$\%$ with collisions and two active regions that are unclassified but consistent with a rising O-ring.

\end{abstract}

\keywords{Sunspots(1653) --- Delta sunspots(1979)}

\section{Introduction} \label{sec:intro}

Sunspot regions that contain both positive and negative magnetic polarity umbrae within 2$\degree$ of each other and within a shared penumbra are defined as $\delta$-spots by \cite{Kunzel:1965}.  They appear in observations as a type of magnetic knot \citep{tanaka:1991} with polarities that do not separate in time in contrast to simpler active regions (ARs). We use the term 'magnetic knot' to refer to the umbrae in an active region that satisfy the criteria of being in a $\delta$-configuration and are participating in the $\delta$-configuration. There can be multiple magnetic knots in a single AR classified as $\delta$ and not all of the umbrae in $\delta$-spots need to be part of a knot as some umbrae are not participating in the $\delta$-configuration.  Due to the fact that $\delta$-spots are disproportionately responsible for the most energetic flares and eruptions during any given solar cycle, they are a topic of interest to solar physics and space weather research \citep{tanaka:1991, shi:1994, sammis:2000, guo:2014}. In contrast to the relatively uncommon, flare-active $\delta$-spots, $\beta$-spots are the most common sunspot group category comprising 64$\%$ of all sunspot groups \citep{Jaeggli:2016}. A simple $\beta$-region has distinct positive and negative polarity umbra that are contained in separate penumbra. 

The scenarios put forward to explain the formation of $\delta$-spots include a highly twisted, kink-unstable flux tube \citep{tanaka:1991, linton:1996, Fan:1999, Takasao:2015, Knizhnik:2018}, convective buffeting of the flux tube, i.e. the Sigma ($\Sigma$-) effect \citep{longcope:1997, longcope:1998}, the multi-segment buoyancy model in which two bipoles emerge at two different locations from a single flux tube \citep{Toriumi:2014}, and the collision of two emerging tubes \citep{Murray:2007,  Jouve:2018}. The twist and the writhe of the flux tube are predicted to be of the same sign if the formation is due to the kink instability and of the opposite sign if the formation is due to the $\Sigma$-effect \citep{linton:1996}.

\begin{figure}[!t]
\begin{center}
\includegraphics[trim=0.005in -0.8in 0.005in 0.005in,clip,width=0.34\textwidth]{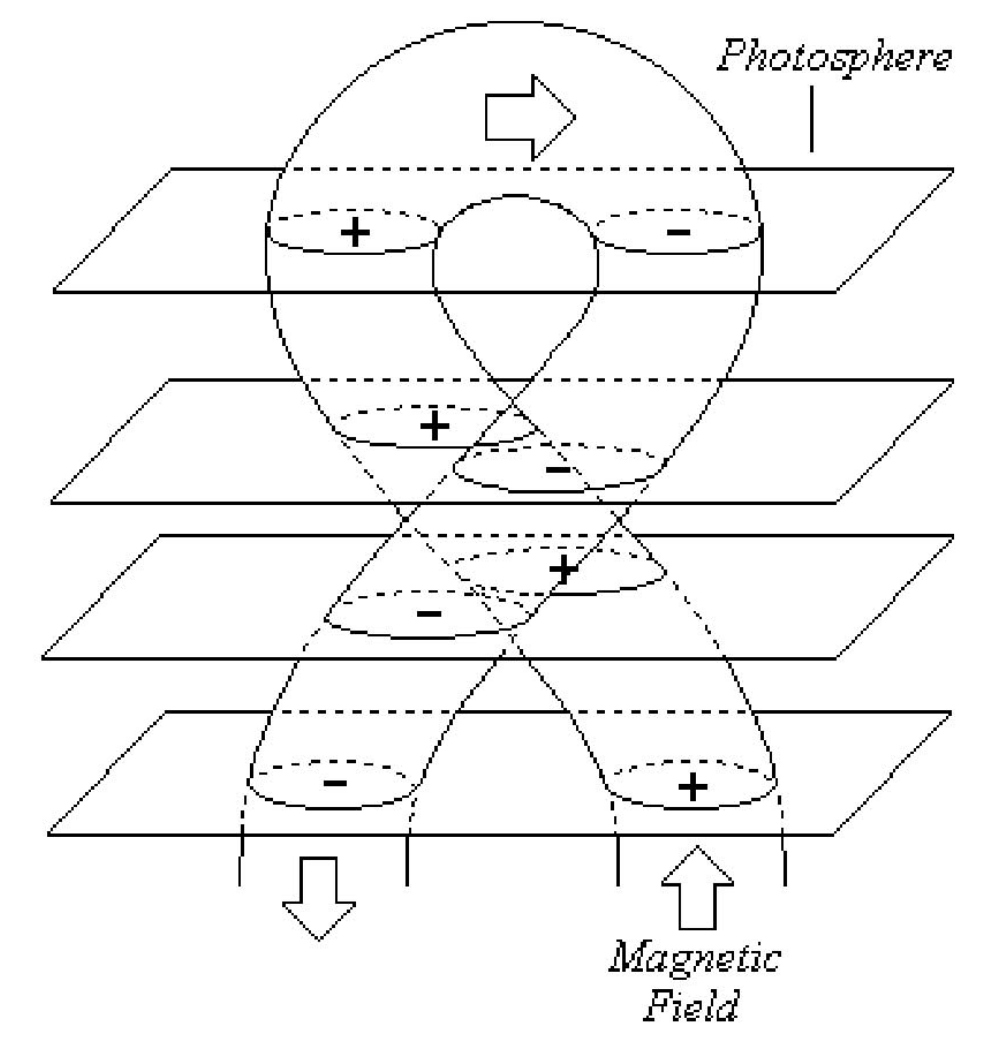}
\put(-110,15){(a)}
\put(60,15){(b)}
\includegraphics[trim=0.005in -0.3in 0.005in 0.005in,clip,width=0.55\textwidth]{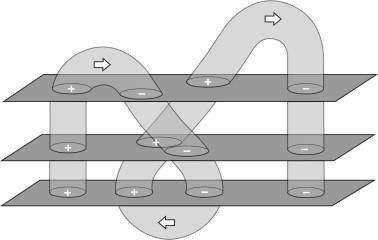}
\end{center}
\begin{center}
\includegraphics[trim=0.0in 0.0in 0.0in 0.0in,clip,width=0.38\textwidth]{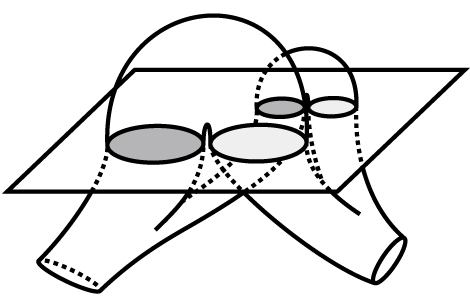}
\put(-100,1){(c)}
\includegraphics[trim=0.0in 0.0in 0.in 0.0in,clip,width=0.38\textwidth]{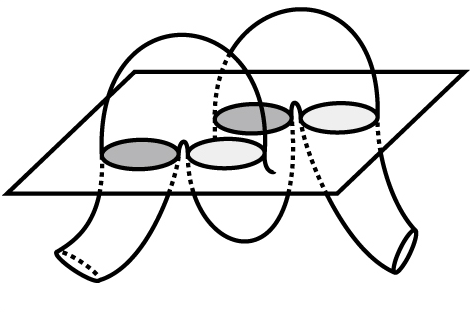}
\put(-90,1){(d)}
\caption{Flux tube geometries that could lead to the formation of $\delta$-spots are shown:  (a) an ``island"-$\delta$  that is the first category of \cite{Zirin:1987}, also known as the ``spot-spot" type by \cite{ToriumiMagProp:2017}, formed by a kink instability acting on a twisted, rising $\Omega$-loop or convective buffeting, image reproduced from \cite{lopez:2000}; (b) an inverted kink instability, reproduced from \cite{poisson:2013}; (bottom row) two multi-segment buoyant configurations with different subsurface connectivity, image reproduced from \cite{ToriumiMagProp:2017} who named them ``spot-satellite" (c) and ``quadrupole" (d).  These last two multi-segment buoyancy configurations represent the second and third category of \cite{Zirin:1987}. The configuration in (a) would appear in a magnetogram as a primarily bipolar magnetic configuration, whereas the other three would present as quadrupolar. These schematics don’t show crucial aspects of flux emergence such as the build-up of flux at the shoulders of the rising loops, moving dipolar features, etc.}
\end{center}
\label{geometry}
\end{figure}

\cite{Zirin:1987} categorized the formation of $\delta$-spots in the following three ways based on decades of observations at Big Bear Solar Observatory. The first type is an AR emerging all at once as a dipole with the dipoles intertwined, often compact with a large umbra and known as an ``island $\delta$".  The second is a $\delta$-spot produced by emergence of satellite spots near large older spots, and often a small opposite polarity umbra is within the older, larger spot's penumbra.  Thirdly, a $\delta$-configuration is formed by the collision between two separate but growing bipoles that emerged nearly simultaneously,  such that the overall area has a quadrupolar configuration and the follower spot of one bipolar region collides with the preceding spot of the other. \cite{ToriumiMagProp:2017} nicely illustrates these three observed formations and the possible subsurface flux tube structures, naming the categories as spot-spot (``island"-$\delta$s), spot-satellite, and quadrupole, see Figure~\ref{geometry} for illustrations of the $\delta$-spot types. Lastly, interacting ARs that supposedly are not connected below the surface can collide and form $\delta$-spots. Using observations to distinguish between the formation mechanisms is challenging as the scenarios shown in Figure 1b and 1d would appear nearly identical unless one could confidently measure current or kinetic helicity to differentiate them. \cite{ToriumiWang:2019} provides a comprehensive review of flare-productive ARs, including a summary of $\delta$-spots.

We simplify the observational categorization of regions to be either: a single flux emergence event bipole (SEEB), a single emergence event quadrupole (SEEQ) or a multiple flux emergence event quadrupole or multipole (MEEQ). Figure 1a represents a SEEB while the other three panels represent SEEQ. 

 $\delta$-spots are classified by observers using the Mount Wilson system via the following method: if a line can easily be drawn between the two polarities, it is classified as a $\beta\delta$ configuration; if the polarity spatial distribution is more complicated and no such neutral line can easily be drawn, it is classified as a $\beta\gamma\delta$ configuration. Most $\delta$-spots are complicated and classified as $\beta\gamma\delta$ with only 16$\%$ of $\delta$s between 1992-2016 being the simpler $\beta\delta$ classification \citep{Jaeggli:2016}. \citet{Jaeggli:2016} reported a variation in AR complexity as a function of solar cycle in that complex ARs (including all groups with a $\delta$ or $\gamma$ classification) comprised a larger percentage of ARs during late solar maximum and the declining cycle phase. \citet{nikbakhsh:2019} confirmed this finding. One interpretation of the cycle trend is that complex ARs (including $\delta$s) are produced by the collision of different magnetic systems when the frequency of flux emergence is high. \citet{Jaeggli:2016} found no significant difference in latitude distribution for complex ARs during the solar cycle. See material in Appendix A for a list of all $\delta$-regions in Cycle 24 and their time-latitude distribution. \cite{sammis:2000} showed that larger flares were well-correlated with more complicated sunspot structure, with $\beta\gamma\delta$-spots hosting the strongest flares.
 
\begin{figure}[b]
\begin{center}
\includegraphics[trim=2.0in 11.86in 2.0in 11.0in,clip,width=0.8\textwidth]{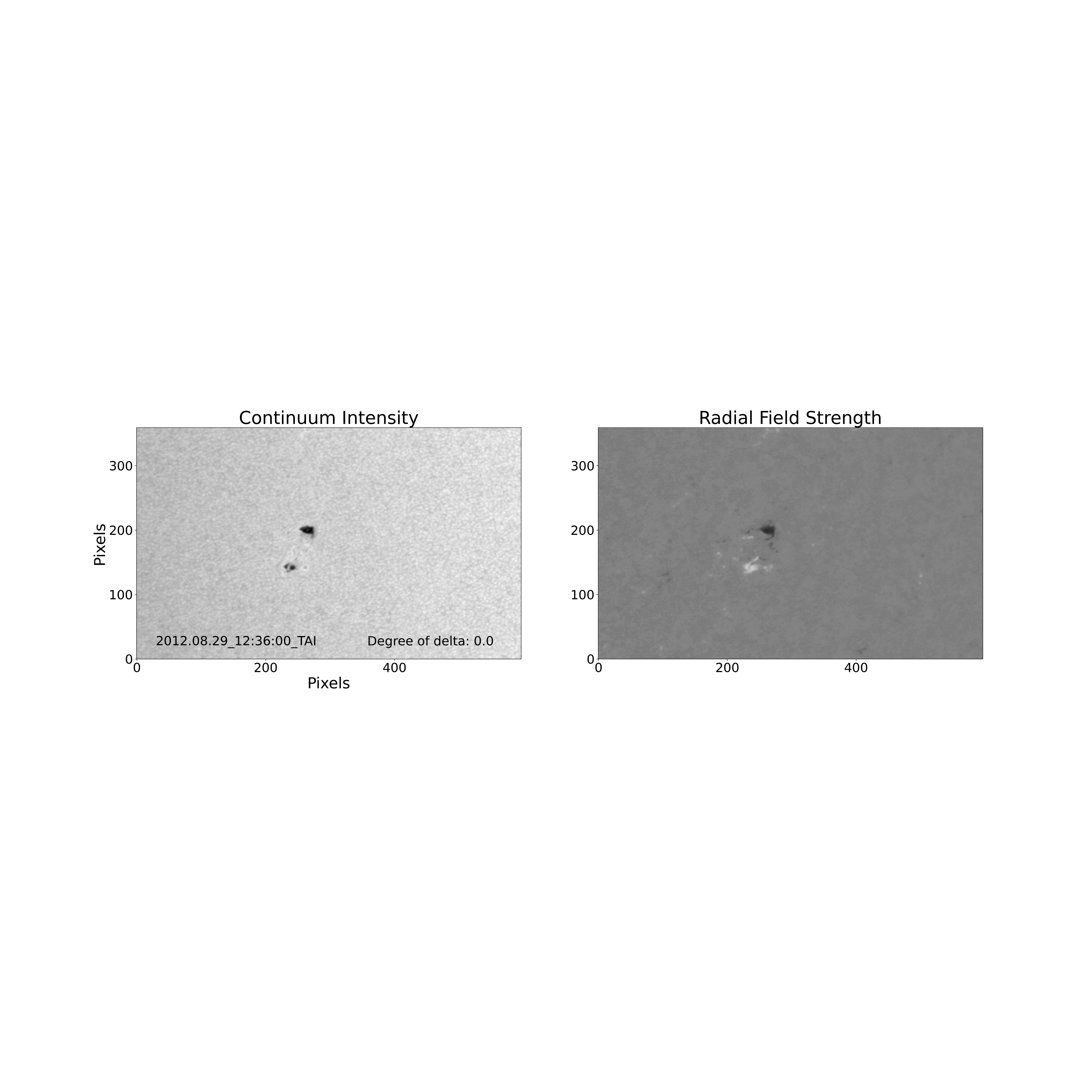}
\includegraphics[trim=2.0in 11.86in 2.0in 11.73in,clip,width=0.8\textwidth]{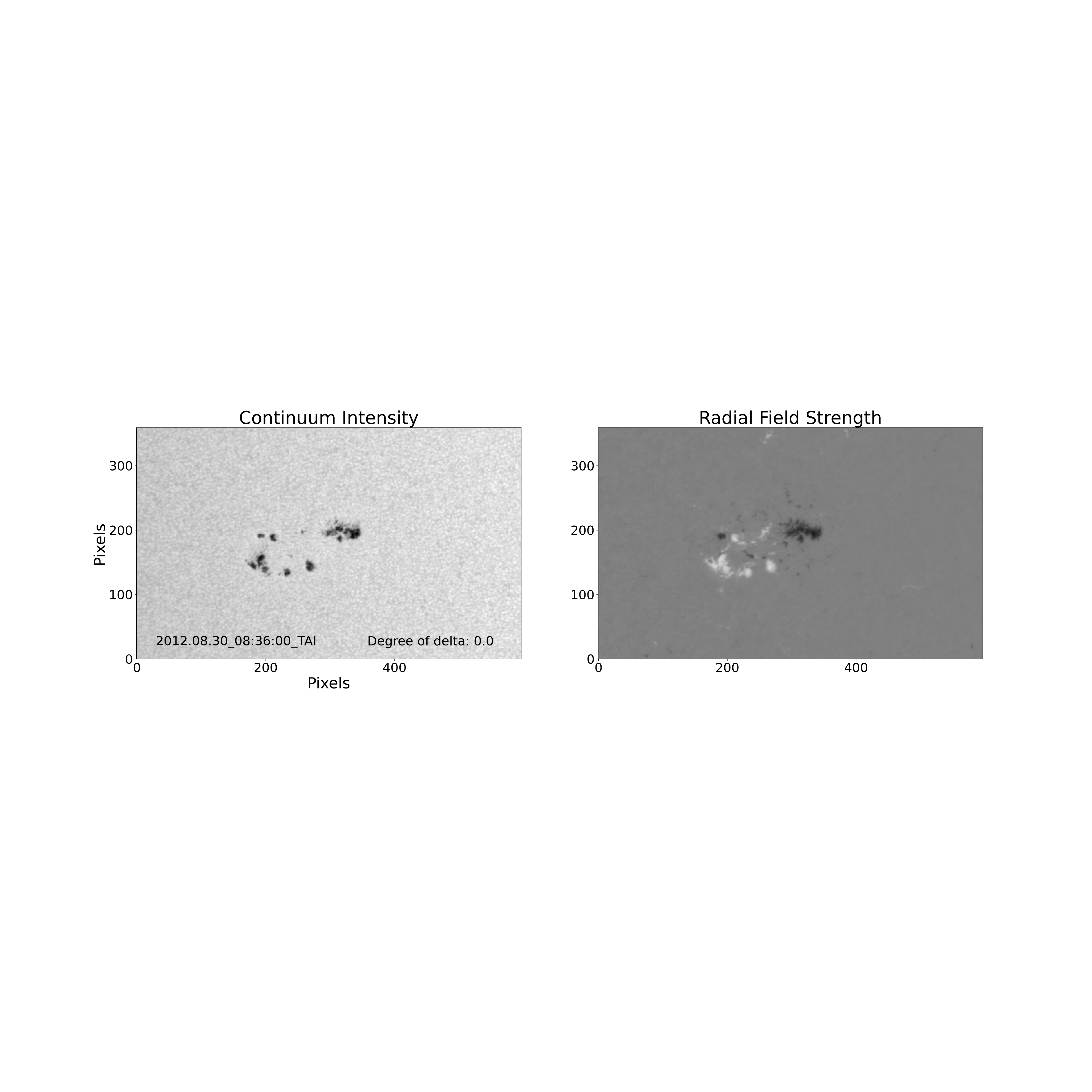}
\includegraphics[trim=2.0in 11.86in 2.0in 11.73in,clip,width=0.8\textwidth]{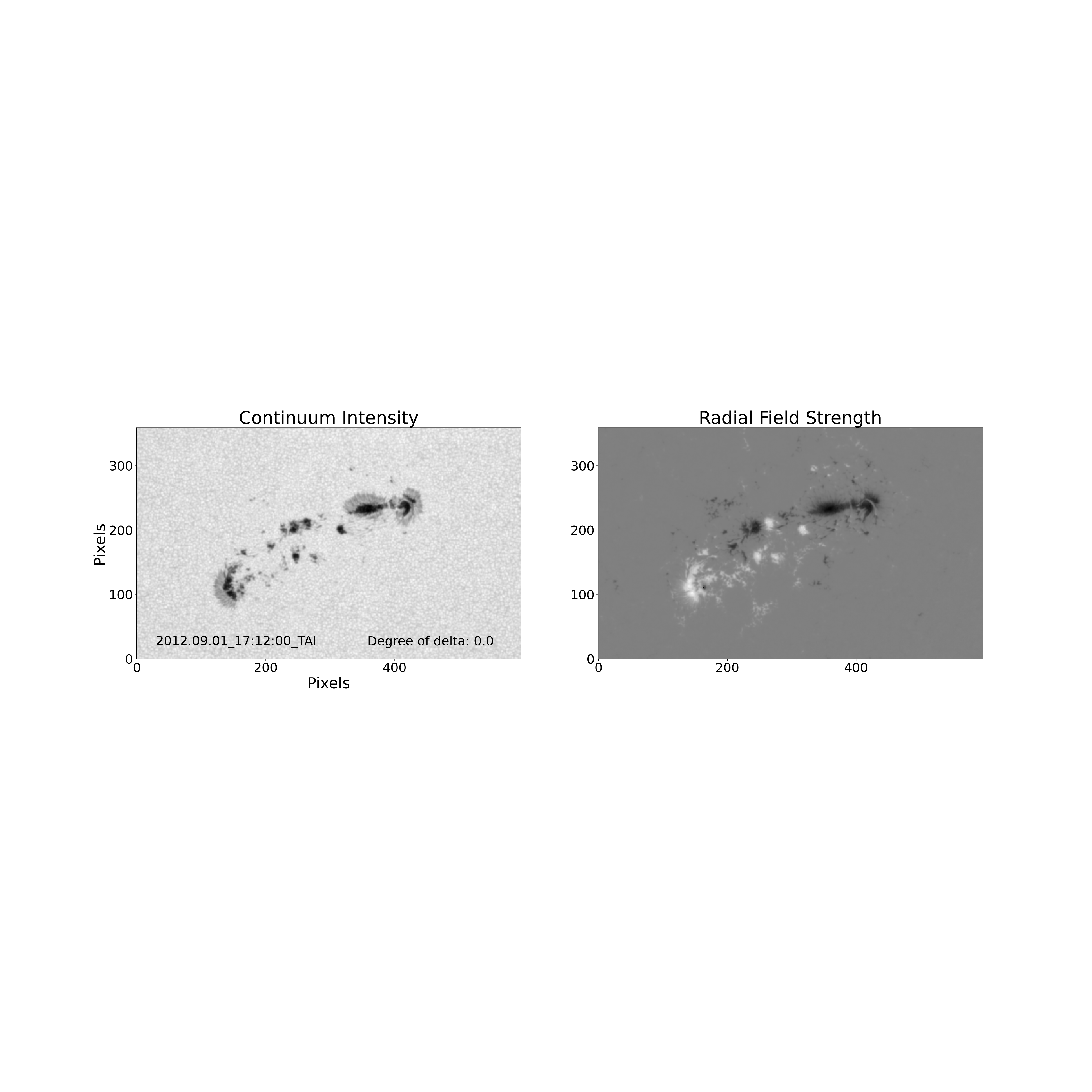}
\includegraphics[trim=2.0in 11.86in 2.0in 11.73in,clip,width=0.8\textwidth]{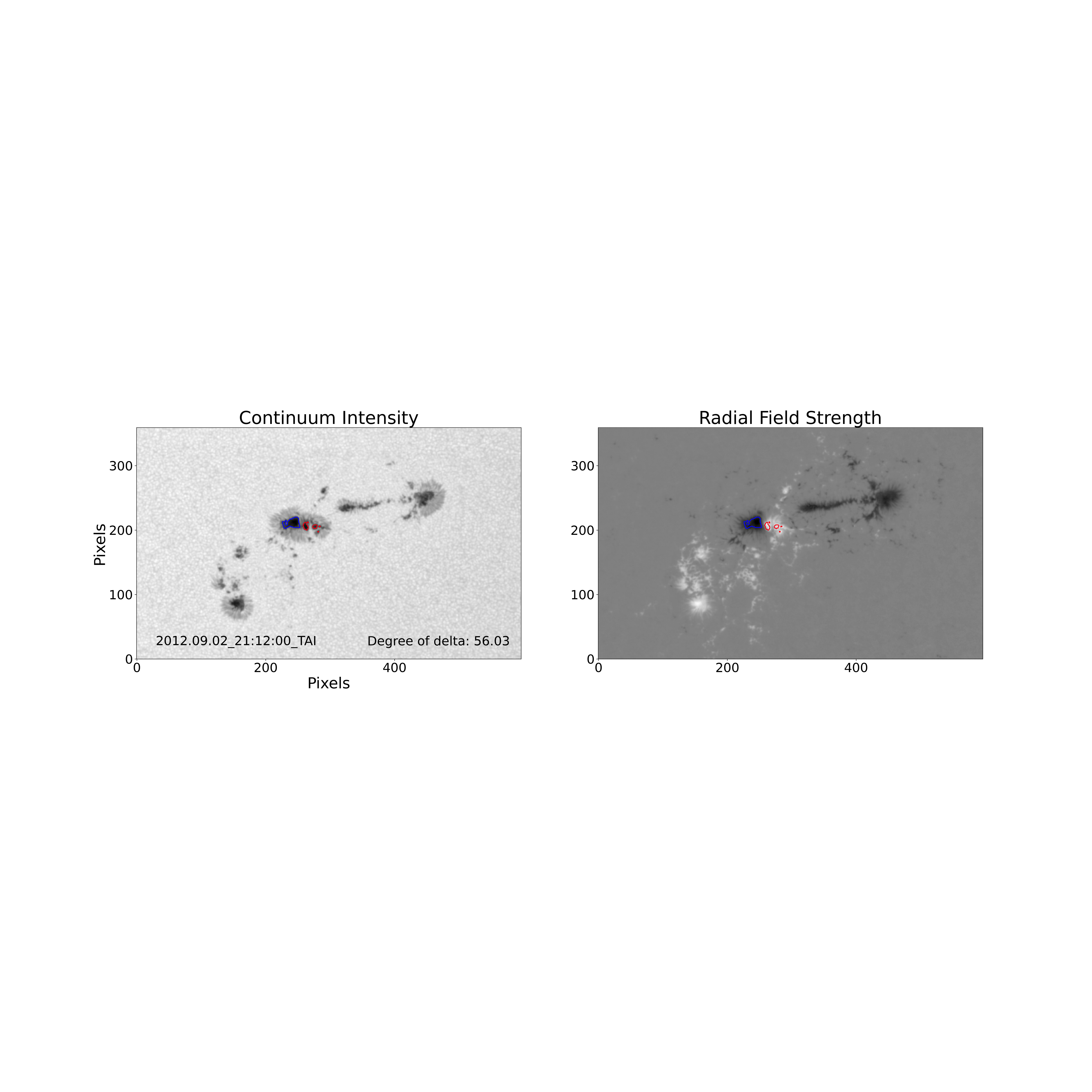}
\includegraphics[trim=2.0in 10.8in 2.0in 11.73in,clip,width=0.8\textwidth]{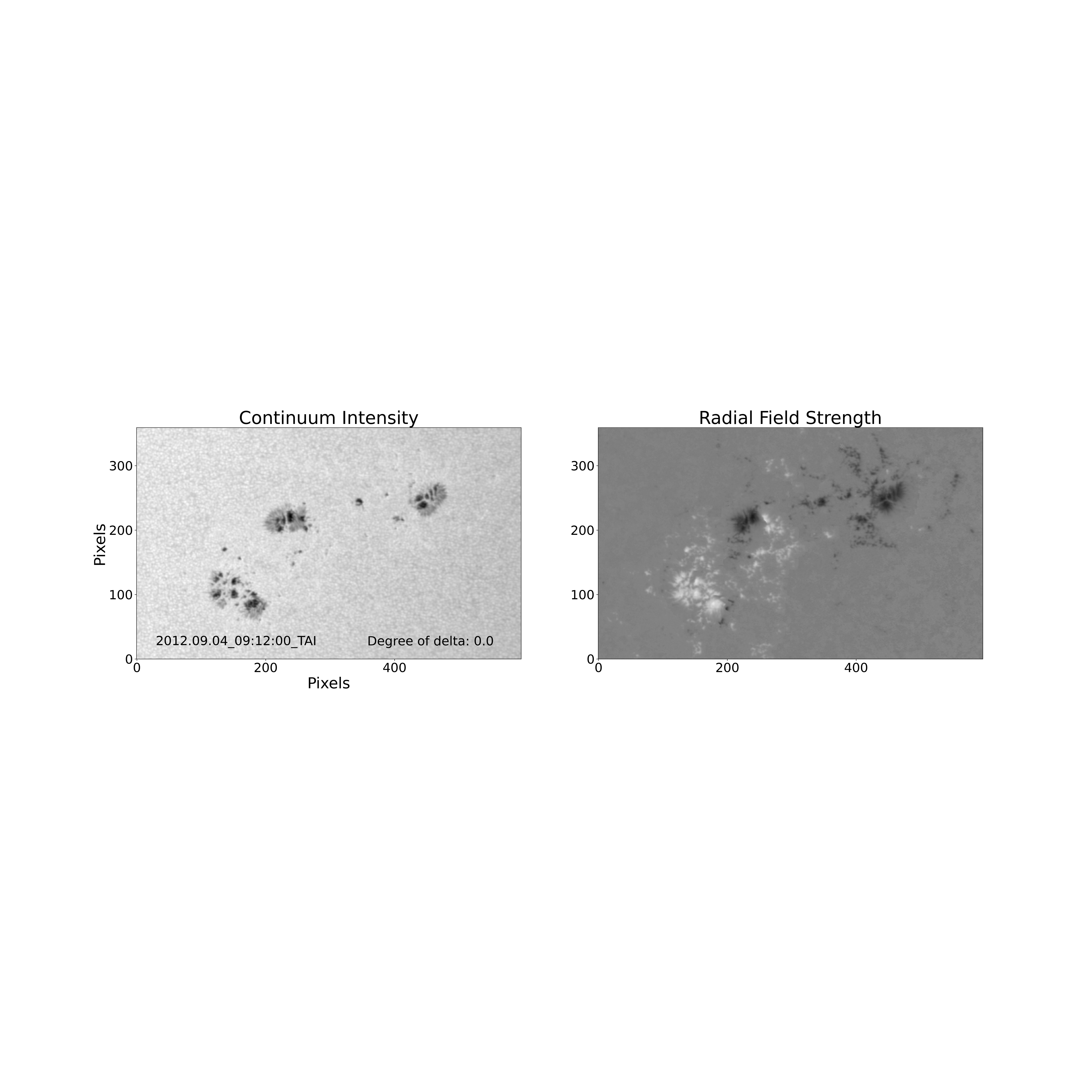}
\end{center}
\caption{NOAA AR 11560 is a $\delta$-spot shown at different times in SDO/HMI continuum intensity (left) and radial magnetic field, B$_r$ (right). The third panel from the top is the time of maximum umbral flux, $\phi_{{\delta}_{max}}$, while the fourth panel from the top is the time of maximum flux in the $\delta$-umbrae only, ${Do\delta}$ of 56$\%$. The umbral regions participating in the $\delta$ configuration are identified by contours on the continuum intensity image with yellow (red) contours identifying the negative (positive) umbra and yellow (red) $\times$ symbols identifying the centroid location for the negative (positive) umbrae in the $\delta$. This is a northern hemispheric AR in which the umbrae participating in the $\delta$-configuration present an anti-Hale tilt but a tilt calculated using all umbrae in the AR presents an anti-Joy tilt, since most northern ARs in Cycle 24 have a positive trailing polarity (white) closer to the pole and this one is closer to the equator. The $\delta$-configuration stays compact and the magnetic knot rotates 70$\degree$ clockwise. The region has a $\delta$-configuration for 29$\%$ of the time observed. It is a SEEQ and its presentation is similar to either the inverted kink configuration shown in Figure~\ref{geometry}b, or the multi-segment buoyancy ``quadrapole" configuration shown in Figure~\ref{geometry}d.}
\label{DoD1}
\end{figure}

If the kink-instability were crucial in forming $\delta$-spots, it is expected that they have certain observational signatures. The tilt angle may deviate significantly from Hale's or Joy's law. Hale’s
law describes that the ordering of the group magnetic polarity in the East-West direction is opposite in the North and South hemispheres for a given sunspot cycle and that this ordering changes from one cycle to the next \citep{Hale:1919}. Joy’s law describes the tendency for the follower spot in a sunspot group to be located more poleward, i.e., at a higher latitude, than the preceder spot and for that poleward displacement to be greater at higher latitudes than near the Equator \citep{Hale:1919}. \cite{Tian:2005} found that 34$\%$ of $\delta$-spots, in a sample of 104, violated Hale's law or Joy's law but the majority of the 104 spots still followed the hemispheric current helicity rule, i.e. the dominance of negative (positive) current helicity in the northern (southern) hemisphere \citep{pevtsov:1995}.  \cite{Knizhnik:2018} modeled kink-stable and kink-unstable flux ropes and found that quantities that can be observed in the photosphere, such as footpoint separation (compactness), rotation, and anti-Hale orientation, all behaved according to the expectations with the kink-unstable regions being more compact, having higher rotation, and having a larger percentage of anti-Hale configurations than flux tubes that were not kink-unstable.  However, \cite{Knizhnik:2018} showed that observable quantities such as the force free parameter alpha, current density and current neutralization ratio do not easily distinguish between the highly twisted or weakly twisted flux ropes. They argued that this could be due to the force free parameter alpha being a poor representation of flux rope twist, or due to twist \citep{fan:2009} and/or current \citep{torok:2014} remaining below the photosphere during emergence \citep{berger:1984}.

Although numerous studies have shown that $\delta$-spots exhibit higher than average compactness and rotation, and are often anti-Hale, these values are derived using the entire AR and thus include regions of the AR that are not part of the $\delta$-configuration. The center of mass and total current in the AR (and other properties) are therefore influenced by parts of the AR that are not related to the umbrae in the $\delta$-configuration. We were motivated to study the $\delta$-spots in a different way.  We quantify the ``degree of $\delta$" (Do$\delta$), i.e., the fraction of umbral flux that is participating in the $\delta$-configuration, the percentage of $\delta$-regions that are anti-Hale or anti-Joy, the flux emergence rate of $\delta$-spots compared to other ARs, the rotation and footpoint separation of the $\delta$-portions of the ARs, and the correlation between Do$\delta$ and flare energy. 

As an example of the Do$\delta$, see Figure~\ref{DoD1}, showing 5 snapshots in chronological order of the emergence and evolution of NOAA 11560, a quadrupole formed by a single flux emergence event. The time of total maximum umbral flux $\phi_{{\delta}_{max}}$ is shown third from the top and the time of maximum umbral flux in the $\delta$-configuration, i.e. $\delta$-umbrae only flux, $\phi_{Do\delta}$, is shown fourth from the top. This illustrates how the $\delta$-configuration does not exist for the full evolution of a given AR and also shows how the tilt of the umbrae participating in the $\delta$-configuration is anti-Hale but the tilt calculated from all the umbrae in the AR is anti-Joy.

\begin{figure}[!t]
\includegraphics[trim=1.5in 0.01in 1.0in .1in,clip,width=.45\textwidth]{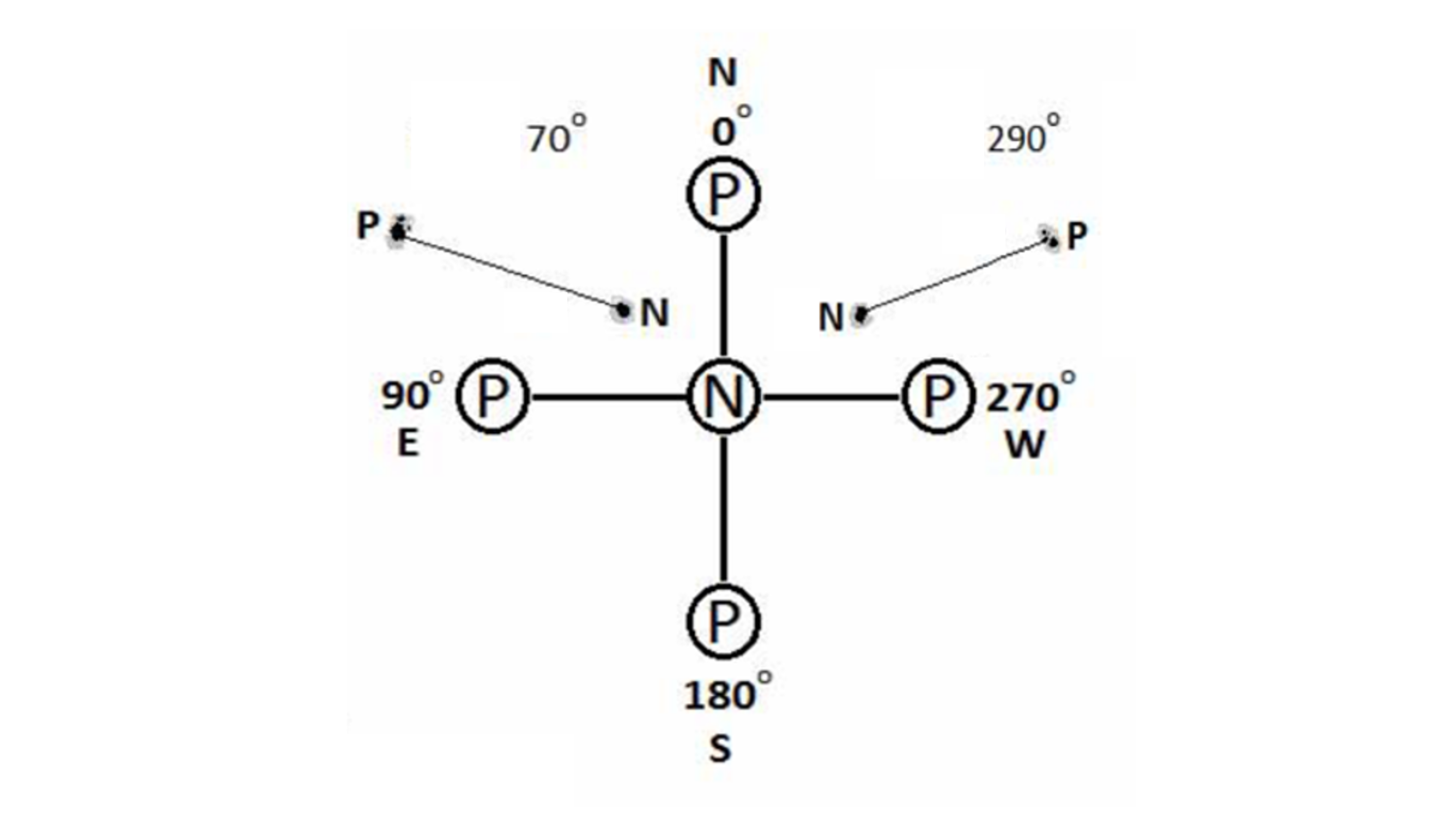}
\includegraphics[trim=0.8in 0.01in 1.5in 0.0in,clip,width=.48\textwidth]{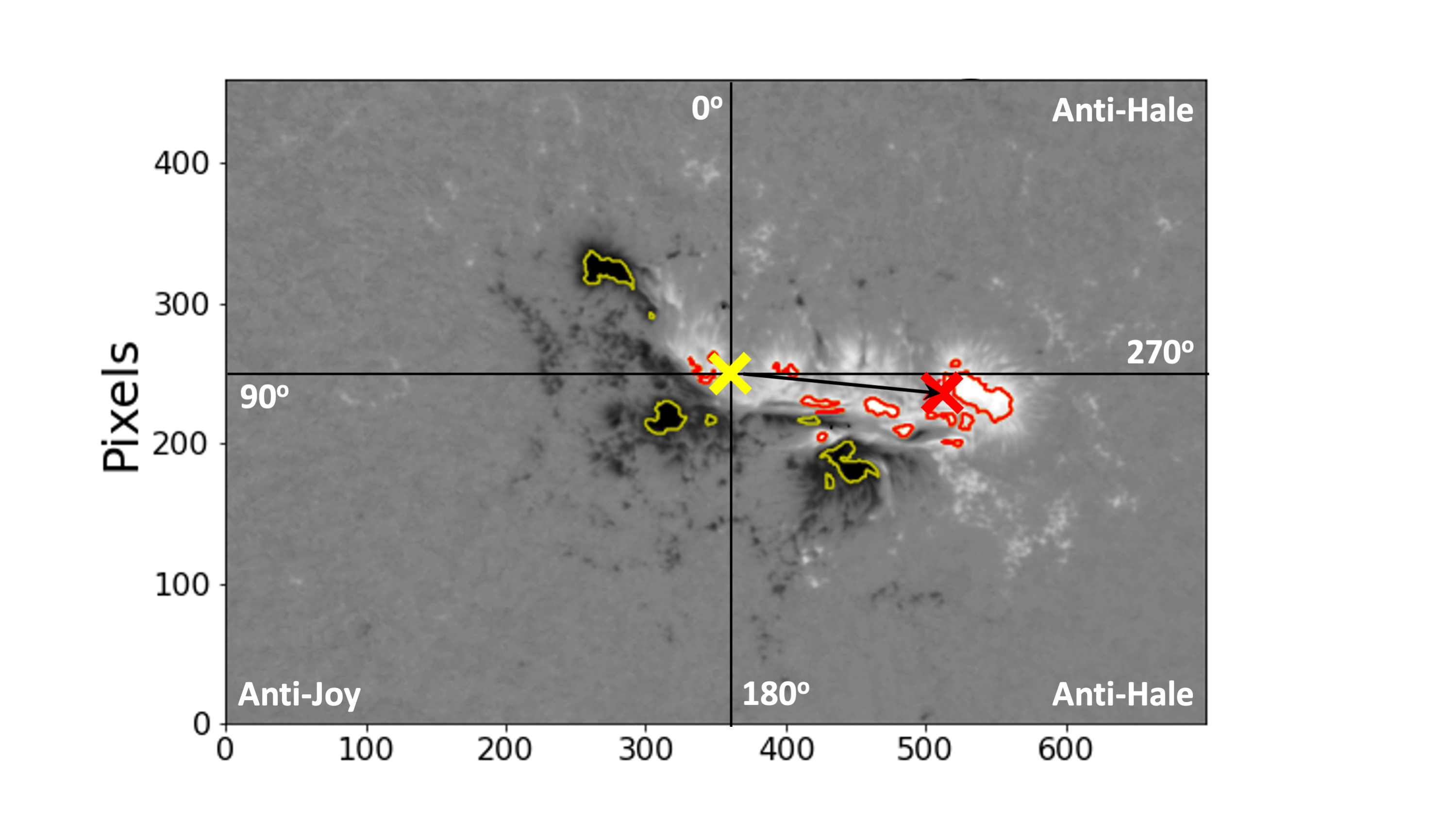}
\caption{Tilt angles are determined with a 0-360$\degree$ range such that the calculated centroid for the flux-weighted negative magnetic polarity ``N" is always at the origin. A 0$\degree$ tilt is an AR in which the``N" and``P" have the same longitude with the "P" at a more northern latitude. Two drawn configurations are shown (left) and can be understood by considering that in Solar Cycle 24, the dominant leading polarity in the northern hemisphere is ``N", so that the example on the left-hand side depicts a common tilt (0-90$\degree$) for most ARs in the northern hemisphere. Historically, using white light images and a limited range of angles (0-90 or 0-180$\degree$), the angle may have been recorded as 20$\degree$ measured from the East-West line, but our angle determination would be 70$\degree$. In contrast, the drawn angle depicted in the 270-359$\degree$ quadrant shows a typical southern hemisphere configuration where the``P" is leading and the``N" is closer to the pole, but instead of a 20$\degree$ angle that may have been reported historically, the angle based on magnetic polarity and a 0-360$\degree$ range is 290$\degree$. While most northern (southern) hemisphere ARs obeying Hale and Joy's law would have a 60-90$\degree$ (270-300$\degree$) values in Cycle 24, $\delta$-spots frequently disobey Hale and Joy's law. At right, the radial magnetic field strength for NOAA 11429 is shown. It is a northern hemispheric AR and has a tilt angle of ~268$\degree$ with the yellow (red) $\times$ representing the negative (positive) centroids as calculated using all umbrae (not only the $\delta$-portion). This is anti-Hale as tilts of 180-359$\degree$ (90-179$\degree$) for northern hemisphere spots in cycle 24 are anti-Hale (anti-Joy), as indicated by the labeling of the magnetogram quadrants for ARs in the northern hemisphere.} 
\label{tilt}
\end{figure}

\section{Data}

The Helioseismic and Magnetic Imager (HMI) instrument on board the Solar Dynamics Observatory (SDO) spacecraft utilizes filtergrams to image the full disk of the Sun at the Fe-I 6173 \text{\AA} absorption line with a pixel size of $\approx$0.5$^{\prime\prime}$ and a 4096$\times$4096 CCD \citep{schou:2012,scherrer:2012}. The filtergram images are recorded for six wavelength positions across the spectral line in a combination of polarization states to acquire the Stokes I, Q, U and V data. The HMI team produces observables of line-of-sight magnetic field values, Doppler velocities, line widths, line depths, and intensities every forty-five seconds while providing vector magnetic field quantities derived from the Very Fast Inversion of the Stokes Vector (VFISV) code every twelve minutes \citep{metcalf:1994,borrero:2011,centeno:2014, hoeksema:2014}. 

During the Solar Cycle 24 years that HMI has been recording data (2010-2019), 132 ARs contained a $\delta$-configuration. The classifications are found in the Solar Region Summary (SRS) files in the NOAA database. The SRS is a joint product of NOAA and the US Air Force (USAF) issued daily providing a detailed daily description of ARs after analysis and compilation of individual reports from the USAF observers using the Solar Optical Observing Network (SOON) that includes ground observatories in Learmonth, Western Australia, San Vito, Italy and Holloman Air Force Base, New Mexico.

For our research, all days that contained a $\delta$ classified region were noted and the AR NOAA numbers and dates were used to find the sunspot data as recorded by HMI. HMI AR data are stored as both full-disk images and as smaller cut-out regions known as Spaceweather HMI Active Region Patches (SHARPs). Each region is assigned a SHARP number and these regions are tracked at the Carrington rotation rate and processed within the HMI pipeline {\citep{bobra:2014}. The SHARPs data contain quantitative parameters describing the regions such as total unsigned flux, flux-weighted longitudinal and latitudinal center of each polarity, etc., that are stored as keywords. The NOAA numbers, SHARP numbers, dates, Mount Wilson categories and maximum flare energy for each $\delta$-spot observed by HMI are found in the Tables in the Appendix. The SHARPs data product used in this research are the hmi.sharp\_cea\_720s, which have a Lambert cylindrical equal area projection.  Specifically, the continuum intensity, $I_c$, and the radial field, $B_r$, segments are analyzed.  We sample the SHARP data every 2-hours in our analysis, although they are available every 12 minutes so could be sampled more frequently if desired. 

Limb darkening is removed for each continuum intensity image using a second-order polynomial reported in \citet{pierce:1977} with a dependence on the center-to-limb angle as follows: $I_c = A + B\times cos(\theta) + C\times (cos(\theta))^2$ where $A=0.36019$, $B=0.90010$, $C=-0.26029$ and the center-to-limb angle $\theta$ is defined as cos($\theta$)=1 at disk center. The limb darkening is important to include to ensure that umbral boundaries are defined in a consistent manner as the region traverses the solar disk. 

Our \textit{small sample} is comprised of 19 $\delta$-spots, and 11 $\beta$-spots, that act as a control group. We also analyze a \textit{large sample} of 120 $\delta$-spots, i.e. all those observed by HMI in Cycle 24 within 55$\degree$ of disk center, calculating more limited quantities. We use the GOES catalogue to identify flares associated with each NOAA numbered AR.  We note every M-class or higher flare, using the sum of these flare X-ray energies as a total flux and the single, highest energy flare as the maximum X-ray flux. 

\section{Analysis}
We characterize the ARs in two distinctly different manners: 1) values characterizing only the $\delta$-portion of the AR umbra at the time of maximum flux, $\phi_{Do\delta}$ and 2) values characterizing the entire AR when the region is at its maximum umbral flux, $\phi_{{\delta}_{max}}$.  We also report on the values of a group of $\beta$-spots at the time of maximum umbral flux, $\phi_{{\beta}_{max}}$, to use as a comparison to the $\delta$-spot values at $\phi_{{\delta}_{max}}$.  

The umbral-penumbral and the penumbral-quiet Sun boundaries are identified using both the $I_c$ and $B_r$ data in an automated manner after limb darkening is removed. Similar to \citet{Norton:2017} and \citet{Verbeeck:2013}, a threshold value is determined from the mean continuum intensity and the standard deviation of the intensity. On average, using an intensity contrast threshold of 85$\%$ for penumbral pixels and 56$\%$ for umbral pixels in comparison to quiet-Sun $I_c$ adequately isolated the umbrae in a given image. The magnetic field values, $B_r$, at the same time are used to determine the polarity of the umbrae. We use a noise threshold of 575 Mx cm$^{-2}$ for the vector field strengths such that anything below this value is considered to be noise.  This noise threshold is higher than the 225 Mx cm$^{-2}$ recommended by \citet{bobra:2014} and does not affect the identification of umbrae that are participaing in the $\delta$-configuration since the umbral field values are always higher than 575 Mx cm$^{-2}$.  However, using the threshold of 575 Mx cm$^{-2}$ does effectively lower values such as the total flux of the region and the flux emergence rate of the region because plage and penumbral fields can fall below 575 Mx cm$^{-2}$.

In order to determine if the $\delta$-criteria is satisfied, and to identify the umbral areas participating in the $\delta$, a binary array the same size as the AR image is created in which penumbral pixel values are assigned as 1 and non-penumbral pixel values as 0. Any contiguous or neighboring penumbral pixels in that array are placed into their own list where each list marks a separate contiguous region of penumbra.  A similar process is performed on the positive and negative umbral regions.  Then, a simple loop through each penumbral region determines if both a positive and negative umbral region are contained within the aforementioned binary subset of penumbra.  If yes, then the umbral regions are classified as in a $\delta$-configuration. We do not implement the criteria that the opposite umbrae are within 2$\degree$ of each other (24 Mm or 33$\arcsec$ at disk center or 66 HMI pixels) as we find that if they are contained within the same penumbra, then the 2$\degree$ criteria is usually satisfied. There are some instances where the opposite umbrae are further than 2$\degree$ apart. 

Tilt angles are determined using a full 360$\degree$ range of values, see Figure~\ref{tilt} for the angle definitions and an example.  First, the flux-weighted centroids are calculated for the negative and positive polarities of the radial magnetic field in the umbrae. Second, the tilt angle is determined.  Third, we determine which of four quadrants the tilt angle resides in.  Only 1 quadrant represents that of compliance with Joy's and Hale's law, 2 quadrants represent anti-Hale (AH), and the remaining one represents anti-Joy (AJ), see Figure~\ref{tilt}.  An angle that is anti-Hale can also be anti-Joy, but we do not assign both AH and AJ to a tilt value, only one or the other. The rotation was initially calculated as the difference between the tilt angle from the final time when it is in a $\delta$-configuration to the first time it is in a $\delta$-configuration. However, we found that within the $\delta$-spot, several knots, i.e. adjacent umbrae in $\delta$-configuration, could exist at different times and if the final tilt referred to a knot that was not present in the beginning of the $\delta$-configuration, then the rotation was not meaningful.  To better characterize the rotation, we isolated individual magnetic knots within each region and noted the direction and amplitude of their rotation via visual inspection of the movies and tilt angle values as a function of time. Positive (negative) rotation values indicate counter-clockwise (clockwise) directions.  The separation is calculated as the distance between the magnetic flux-weighted centroids of the negative and positive polarities. 

The Do$\delta$ is determined by summing the total, absolute value of the magnetic flux of the umbrae participating in the $\delta$-configuration and calculating the fraction it represents of all the flux in the umbrae of the AR.  Figure~\ref{DoD2} shows four ARs, NOAA 11302, 11429, 12158 and 12673 in $I_{c}$ and $B_r$ at various times with a Do$\delta$ ranging from 17-100$\%$.  The negative and positive flux are not required to be balanced.  

Similar to \cite{Norton:2017}, we calculate the flux emergence rate as the total flux that emerged from a time nearest 10$\%$ and through to 90$\%$ of maximum umbral flux divided by the number of hours in which that occurred. Sometimes, we do not sample the entire emergence so we simply calculate this quantity from the nearest times to 10 or 90$\%$, or for regions whose emergence we do not capture at all, we simply report a rate of zero.  
\begin{figure}[b]
\begin{center}
\includegraphics[trim=1.5in 0.0in 6.0in 0.0in,clip,width=0.97\textwidth]{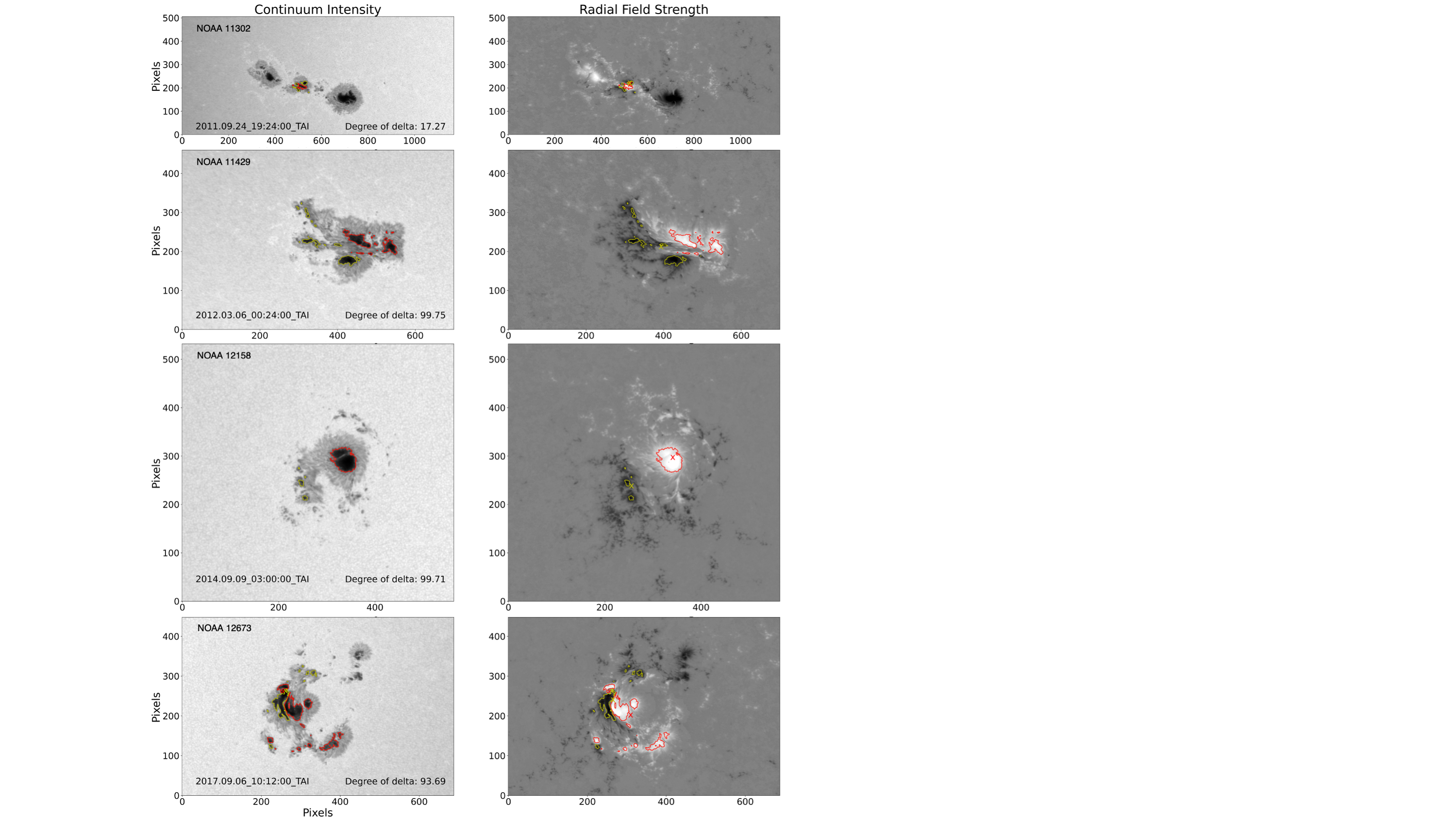}
\end{center}
\caption{Snapshots of NOAA ARs (HARP) 11302 (892), 11429 (1449), 12158 (4536), 12673 (7115) are shown from top to bottom in continuum intensity (left) and radial magnetic field, B$_r$, (right). The umbral regions participating in the $\delta$-configuration are identified by yellow (red) contours on the continuum intensity image outlining the negative (positive) umbra. The corresponding Do$\delta$ at each time is written in the continuum intensity image.}
\label{DoD2}
\end{figure}

Current helicity is a quantity that describes the structure of ARs. We use current helicity to examine the so-called hemispheric preference of the current helicity sign (hereafter referred to as simply the hemispheric rule)  \citep{abramenko1996, bao:1998} for our sample of ARs. The current helicity is computed from the HMI vector field data by $H_z^c = \sum [B_z (\bigtriangledown \times \mathbf{B})_z]dS$ where dS is a differential surface area \citep{abramenko1996,leka2003}. In determining current helicity, we do not isolate the $\delta$-region apart from the entire AR, nor do we examine the current helicity values as a function of time correlated to the maximum Do$\delta$.  We only use the current helicity value as determined within 12 hours of the time the AR crosses the central meridian.

We categorize the $\delta$-spots as either single or multiple flux emergence events (SEE or MEE), and as bipoles (B) or quadrupoles (Q).  A multiple flux emergence event is one in which new magnetic flux emerges at least 48 hours after the initial flux emergence and increases the flux by at least 30$\%$.  These categories are then known as single emergence event bipole (SEEB), single emergence event quadrupole (SEEQ), and multiple emergence event quadrupole or multipole (MEEQ). 

After this categorization, we have a second, less-confident categorization of the regions comparing the observed parameters with signatures consistent with the formation mechanisms of the kink instability or $\Sigma$-effect, multi-segment buoyancy events or collisions by using the parameters of tilt (AH or AJ configuration), Do$\delta$, rotation of the magnetic knot, and the number of flux emergence events observed. If there are multiple flux emergence events separated by at least 48 hours, the region is considered formed via collision of flux systems.  If the region is a bipole with an AH or AJ tilt and a Do$\delta$ higher than 50$\%$, it is categorized as being consistent with a kink instability or $\Sigma$-effect.  Single emergence events with quadrupolar configurations are consistent with both a multi-segment buoyancy or an inverted kink configuration.  We distinguish between these two by arguing that a higher rotation of the central umbrae in $\delta$-configuration ($>$90$\degree$) makes it more likely to be an inverted kink configuration.  The criteria used to categorize a ``spot satellite" was that one of the bipoles in the quadrupole contained much less flux than the other.

\cite{linton:1996} predicted that positive current helicity should correlate with positive (counter clockwise) rotation of the emerging magnetic knot, and vice versa, for the kink instability. We are not using this criteria within this paper, and therefore cannot definitively identify that the $\delta$-spot formation is caused by the kink instability, because we have not yet calculated the current helicity for the isolated magnetic knot; we only have current helicity values for the entire AR.  We hope to address this issue in a future publication.

In summary, for our analysis, we collate a small sample (19 $\delta$-spots and 11 $\beta$-spots that act as a control group), then calculate several instantaneous quantities of the ARs as well as some values derived over the lifetime of the AR.  For the $\delta$-spots, we first identify the portion of umbrae participating in the $\delta$-configuration and report on the following quantities relating to the $\delta$-umbrae only portion at  the time the region is at the maximum $\phi_{Do\delta}$: if the tilt of the bipole related to the $\delta$-umbrae only is anti-Hale or anti-Joy, the Do$\delta$, the unsigned umbral flux participating in the $\delta$ at the time of maximum $\phi_{Do\delta}$, and the polarity separation at the time of maximum $\phi_{Do\delta}$. In addition to these instantaneous values, we report on the total rotation of the tilt angle, the change in separation distance, and the lifetime of the $\delta$-umbrae only portion, and total flare energy.

Next, we analyze the same sample of 19 $\delta$-spots, but this time include all the umbrae, irregardless of participation in the $\delta$ and report the following quantities at time of maximum total umbral magnetic flux, $\phi_{{\delta}_{max}}$: if the tilt of the bipole as determined by all AR umbrae is anti-Hale or anti-Joy, the Do$\delta$, the unsigned total flux of the AR above the threshold, the signed flux emergence rate of all the flux above the threshold in the region from the time the flux is at 10 - 90$\%$ of $\phi_{{\delta}_{max}}$, the polarity separation. In addition to these instantaneous values, we report on the total rotation of the tilt angle over the lifetime, the change in separation distance over the lifetime, the lifetime of the umbrae of the AR, and total flare energy. 
The $\beta$ active regions are analyzed in the same manner as the $\delta$-spots analyzed at maximum umbral flux.   We then use the observed emergence and measured parameters to categorize the ARs into regions whose behavior is consistent with certain formation mechanisms.  

For a snapshot of the temporal analysis for the AR in a $\delta$-configuration, see Figure~5 of NOAA 11158, with the $I_c$ and $B_r$ snapshots and the quantities as described above for analyzing the $\delta$-umbrae only, i.e. the magnetic knot, plotted as a function of time.  The vertical, dashed black line indicates the time in the sequence corresponding to the grayscale images in the top panels.  Figure~6 is a snapshot of the same AR (NOAA 11158) but showing the result of the analysis of the AR taking into account all umbrae, not just the umbrae in the $\delta$-configuration.  The contours are drawn around all umbrae, and the corresponding umbral flux, region flux, tilt and polarity separation are shown. 

Movies are available online for the small sample of 19 $\delta$ and 11 $\beta$ spots included in this paper. Note that we do not employ any smoothing. For the large sample of 120 $\delta$-spots, i.e. all those observed by HMI in Cycle 24 within 55$\degree$ of disk center, we calculate more limited quantities of maximum Do$\delta$, maximum total umbral flux, $\phi_{{\delta}_{max}}$, maximum flare energy, total flare energy and anti-Hale or anti-Joy tilts. 

\section{Results}

Values characterizing the $\delta$-umbrae only portion of the AR (the knot) at its maximum $\phi_{Do\delta}$ are found in Table 1 while values characterizing the entire AR when at its maximum umbral flux, $\phi_{{\delta}_{max}}$, are found in Table 2. The similar characteristics of $\beta$-spots at maximum umbral flux, $\phi_{{\beta}_{max}}$ are shown in Table 3. On the left side of the tables, columns 3-7, are instantaneous values determined at the maximum flux times, while columns 8-11 are time-derived values.

For a summary of whether the tilts are anti-Hale or anti-Joy, see Tables 1-2 or Figure~\ref{anti-hale}. 
\begin{itemize}
    \item[$\ast$] 72$\%$ of the small sample of $\delta$-spots are anti-Hale (44$\%$) or anti-Joy (28$\%$) when these tilts are determined from only the umbrae participating in the $\delta$, 
    \item[$\ast$] 53$\%$ of the small sample of $\delta$-spots are anti-Hale (16$\%$) or anti-Joy (37$\%$) at maximum umbral flux using the traditional method of including all umbrae in the AR to determine the tilt angle, 
    \item[$\ast$]and these values contrast with only 10$\%$ in the $\beta$ group (all Anti-Joy). 
\end{itemize}
The large sample confirms these percentages with 74$\%$ of the $\delta$-umbrae only spots having tilts that are anti-Hale or anti-Joy. This percentage drops down to 50$\%$ of the tilts being anti-Hale or anti-Joy when calculating the tilt using all umbrae in the AR at $\phi_{{\delta}_{max}}$, see Figure~\ref{anti-hale}.

On average, the amount of umbral flux participating in the $\delta$-configuration, the Do$\delta$, was found to be a maximum of 72$\%$ $\pm$ 19$\%$, see Table 1. When calculating the Do$\delta$ at maximum total umbral flux, the value was lower at 41$\%$ $\pm$ 33$\%$, see Table 2.  The umbrae were in a $\delta$-configuration 55$\%$ of the time they were observed.  

The $\delta$-spots show more rotation, and less footpoint separation than the control group.  The total, average change in rotation of the $\delta$-umbrae only portion of the ARs was 62$\degree$, while it was 23$\degree$ when considering all umbrae in the $\delta$, see Tables 1 and 2.  The $\beta$-regions' total change in rotation was, on average, 9$\degree$, see Table 3.   The polarity separation of the $\delta$-only portions of the umbrae were on average, 32.1 Mm and on average, converged over time, see Table 1.  In contrast, when examining all umbrae in the $\delta$-spot, the average polarity separation was 61.1 Mm and on average separated 9.18 Mm over the lifetime of the region, see Table 2.  The $\beta$-spots had an average separation of 66.10 Mm, see Table 3, with an overall change in the separation being a divergence of 31.7 Mm.

\begin{figure}[!t]
\begin{center}
\includegraphics[trim=0.3in 3.0in 0.3in 3.0in,clip,width=.85\textwidth]{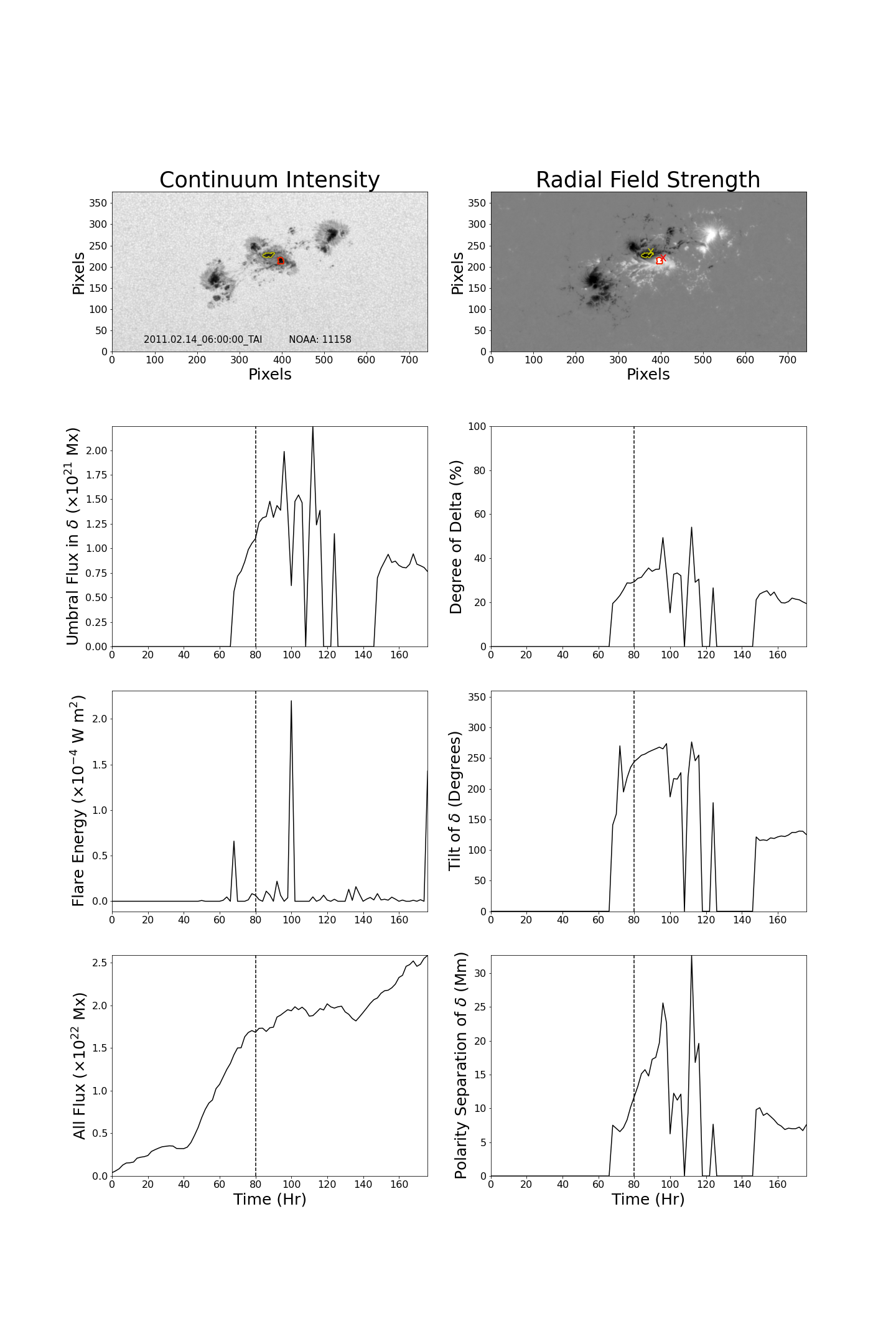}
\caption{This $\delta$-spot is analyzed using the umbral flux participating in the $\delta$-configuration, $\phi_{Do\delta}$, i.e. the umbral flux in the 'knot'.  NOAA AR 11158 (HARP 377) is shown using $I_c$ (left) and $B_r$ (right). The still frame image is from 2011.02.14 at 06:00:00 UT while the duration of the movie is 176 hours from 2011.02.10 22:00:00 until 2011.02.18 06:00:00 UT. The negative (positive) umbral areas participating in the $\delta$-configuration are identified by contours on the $I_c$ image in yellow (red). The flux-weighted centroids of the knot are indicated by yellow and red $\times$ symbols on the $B_r$ image. Below the images are plots as a function of time of the following unsmoothed quantities: the umbral flux in $\delta$-configuration, $\phi_{Do\delta}$, the Do$\delta$ which is the umbral flux in $\delta$ divided by the total umbral flux of the region, flare energy of the entire region, tilt angle of the polarities of the knot, total unsigned flux of the region summed from pixels whose values are above 575 Mx cm$^{-2}$, and the polarity separation of the knot. The dashed, vertical line on the graphs indicates the time corresponding to the grayscale images in the top panels.}
\end{center}
\label{Snapshot}
\end{figure}

\begin{figure}[!t]
\begin{center}
\includegraphics[trim=0.0in 1.4in 0.0in 2.4in,clip,width=.9\textwidth]{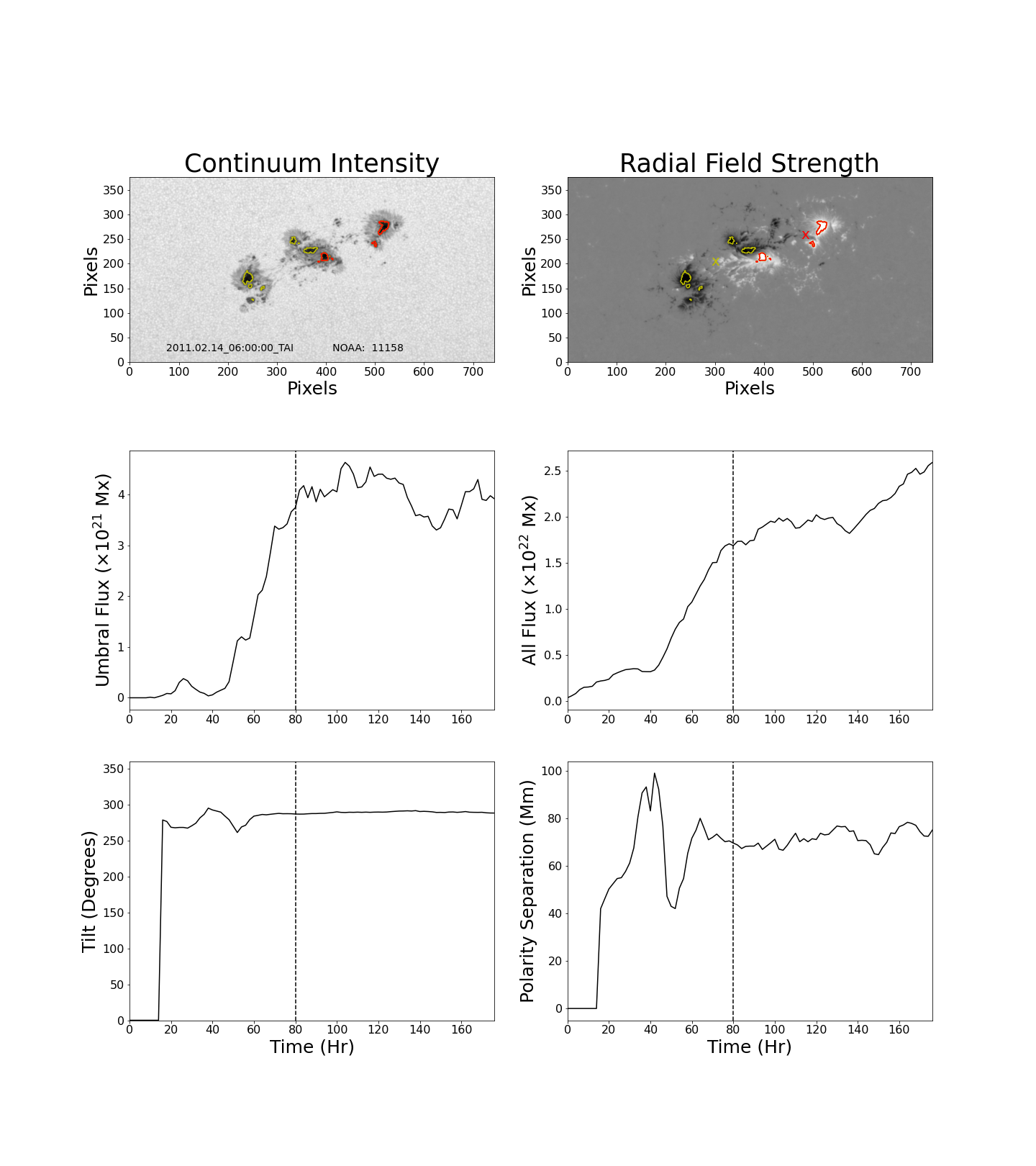}
\caption{The same $\delta$-spot, NOAA 11158, as in the previous figure but analyzed using all umbrae of the AR. It is shown using $I_c$ (left) and $B_r$ (right). The still frame image is from 2011.02.14 at 06:00:00 UT while the duration of the movie is 176 hours from 2011.02.10 22:00:00 until 2011.02.18 06:00:00 UT. All of the negative (positive) umbral areas are identified by contours on the $I_c$ image in yellow (red). The flux-weighted centroids of all umbral polarities are indicated by yellow and red $\times$ symbols on the $B_r$ image. Below the images are plots as a function of time of the following unsmoothed quantities: umbral flux in the AR, the total unsigned flux above the threshold of the region, tilt angle as defined by all umbrae, and the polarity separation as defined by the flux weighted centroids determined using all umbrae. The dashed, vertical line on the graphs indicates the time corresponding to the grayscale images in the top panels.}
\end{center}
\label{Snapshot-Control}
\end{figure}

The flux emergence rates, $\dot{\phi}_{{\delta}_{max}},$ of the $\delta$-spots, whose values are shown in Table 2, were determined by finding the maximum unsigned flux value above the threshold of all flux in the region at the time of $\phi_{{\delta}_{max}}$, then subtracting the flux values at 10$\%$ of that maximum flux from 90$\%$ of the maximum flux, and dividing through by the time elapsed. This value was divided by two in order to report the signed flux emergence rate.  The average flux emergence rate for the full AR of the $\delta$-spots was 10.41 $\times$ 10$^{19}$ Mx hr$^{-1}$ and it was 5.45 $\times$ 10$^{19}$ Mx hr$^{-1}$ for the $\beta$-spots.  
Putting the emergence rates into perspective, see Figure~8 where we overlay a fit (dashed line) reported by
\cite{otsuji:PASJ} of a power-law relationship from Hinode observations where maximum flux emergence
rates were dependent on maximum flux. Also plotted is the fit (solid line) reported by \cite{Norton:2017} of an average flux emergence rate scaling with total signed maximum flux from HMI observations. A value reported for the maximum flux emergence rate by \cite{Toriumi:2014} is plotted as a green triangle. Two simulations of AR flux emergence from \cite{rempel:2014, chen:2017} are also shown using red symbols, see Section 5 for more details and discussion.

\begin{figure}[!ht]
\includegraphics[trim=0.9in 0.2in 1.2in 0.0in,clip,width=0.28\textwidth]{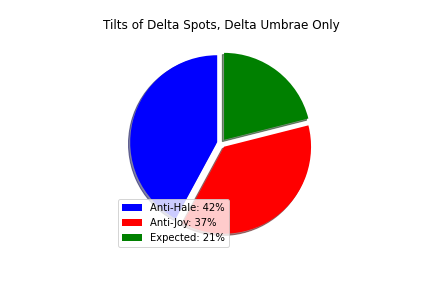}
\includegraphics[trim=0.9in 0.2in 1.2in 0.0in,clip,width=0.28\textwidth]{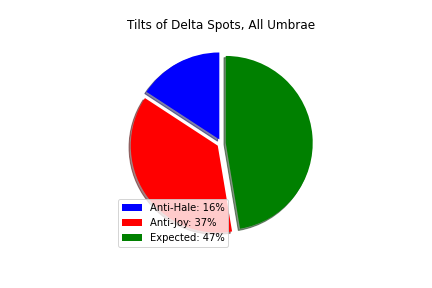}
\includegraphics[trim=0.9in 0.2in 1.2in 0.0in,clip,width=0.28\textwidth]{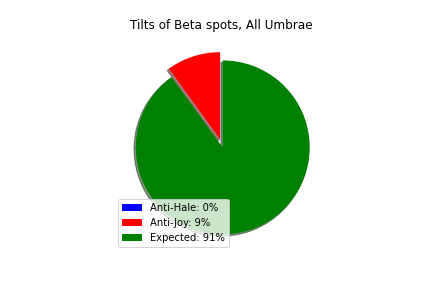}\\
\includegraphics[trim=0.9in 0.2in 1.2in 0.0in,clip,width=0.28\textwidth]{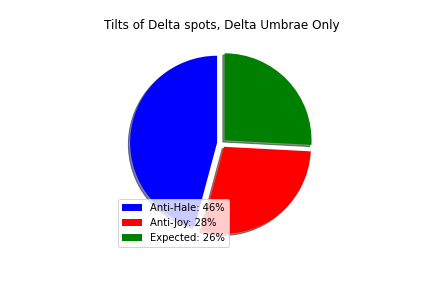}
\includegraphics[trim=0.9in 0.2in 1.2in 0.0in,clip,width=0.28\textwidth]{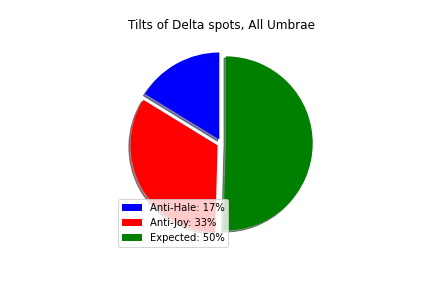}
\caption{\textbf{Top row: small sample.  Bottom row: large sample.} The AR tilts are binned by whether they are in the range of expected values (meaning they obey the Hale's and Joy's law), or whether they have anti-Hale or anti-Joy tilt angles. The pie charts from left to right on the top row represent the tilts of the small sample of 19 $\delta$-spots and 11 $\beta$-spots for only the umbrae involved in the $\delta$-configuration at the time of maximum Do$\delta$, then all umbrae in the $\delta$ AR at the time of maximum umbral flux, and all umbrae for the $\beta$-spots at the time of maximum umbral flux. The tilts from the large sample of 120 $\delta$-spots are shown in the bottom row.  Plotted left to right are tilts determined for umbrae involved in the $\delta$-configuration at the time of maximum Do$\delta$, then all umbrae in that region at the time of maximum umbral flux. One can see that the small and large samples are equivalent in that their pie charts differ only in a few percent for the $\delta$-only analysis. Movies for all regions are found online. }
\label{anti-hale}
\end{figure}

In Figure~\ref{flares}, top panel, the maximum flux in $\delta$, $\phi_{Do\delta}$ (black symbols) for the large sample of 120 $\delta$-spots is plotted against the single, maximum flare energy produced by that AR.  Also plotted in the top panel is the relationship between the maximum umbral flux in the $\delta$-spot, $\phi_{{\delta}_{max}}$ (red symbols), and the single, maximum flare energy. The fit for the relationship between  $\phi_{Do\delta}$ and the single maximum flare energy emitted by that region is $E_{FlareMax}\propto(\phi_{Do\delta})^{0.44}$. The fit for $\phi_{{\delta}_{max}}$ and the total flare energy emitted by that region is $E_{FlareMax}\propto(\phi_{Do\delta})^{0.59}$.  The Pearson correlation coefficients, known as the r-values, are 0.37 (0.39) for the  $\phi_{Do\delta}$ ($\phi_{{\delta}_{max}}$) fits in the top panel with p-values, the probability that the null hypothesis is true, of 10$^{-5}$ (10$^{-5}$).

 The same quantities, $\phi_{Do\delta}$ and $\phi_{{\delta}_{max}}$, are shown in the lower panel of Figure~\ref{flares} as a scatter plot against total X-ray flare energy  associated with each AR during the times observed. The fits for these have exponents of c = 0.54 (0.76), corresponding r-values of 0.47 (0.52) and p-values of 10$^{-8}$ (10$^{-9}$). All fits provide a positive correlation but the correlation between maximum umbral flux, $\phi_{{\delta}_{max}}$ and total X-ray flux, $E_{FlareTot}$ has the highest r-value of 0.52.
 
We examine the hemispheric rule by studying the current helicity for these $\delta$-spots. The current helicity is plotted as a function of latitude and year in Figure 10 with red (blue) indicating negative (positive) current helicity which is expected to dominate in the northern (southern) hemisphere.  The $\delta$-spots are observed to follow the hemispheric rule 64$\%$ of the time, with 71$\%$ in the northern hemisphere and 57$\%$ in the southern hemisphere. We separated out the regions that were anti-Hale and anti-Joy to plot their helicity as a function of hemisphere and time, see the bottom plot in Figure 10, and find that 74$\%$ of anti-Hale $\delta$-regions exhibit the hemispheric current helicity preference (83$\%$ in N and 63$\%$ in S) as do 60$\%$ of the anti-Joy regions (65$\%$ in N and 55$\%$ in S).

We categorize the $\delta$-regions in regards to how many flux emergence events occur and whether they are bipolar or quadrupolar to find 58$\%$ are SEEQ, 26$\%$ are SEEB and 16$\%$ are MEEQ, see Table 4. By definition, the $\beta$-regions are bipoles

We then performed a second categorization using the measured parameters and our understanding of the physical mechanisms responsible for forming $\delta$-spots, see Figure 1, to find that the ARs are consistent with the following configurations:  kink instability or $\Sigma$-effect, inverted kink instability, multi-segment buoyancy configurations such as spot-satellite or quadrupoles, and colliding active regions. We use the AH, AJ, Do$\delta$ and $\Delta$Rot values as determined from the $\delta$-umbrae only, or the `knot', see Table 1.  We note if the region is a quadrupole.  We also note if there were multiple flux emergence events or only a single one.  If there are multiple flux emergence events, defined as emergence events separated by more than 48 hours, the region is immediately classified as being consistent with colliding, or interacting, active regions. If it is a bipole with AH or AJ tilt and a Do$\delta$ higher than 50$\%$, it is classified as being consistent with a kink instability or $\Sigma$-effect. Single emergent events with quadrupolar configurations could be caused by either a multi-segment buoyancy ``quadrupole" or inverted kink configuration. We cannot truly distinguish between the two, but speculate that a higher rotation of the central umbrae in $\delta$-configuration ($>$90$\degree$) makes it more likely to be an inverted kink configuration because the rotation is consistent with the signature of the writhe of the kink rising through the photosphere, as opposed to the follower spot from one region and the leader spot of the other simply joining together because they are joined below the surface.  Some of these ARs, such as 11302 and 12443, do not display AH or AJ. 12158 is compact with an AH tilt with a high Do$\delta$ so it appears consistent with a kink instability but has very little rotation. 12715 is unusual because it is not AH nor AJ, shows very little rotation, emerges with signatures consistent with an arch rising but then the magnetic polarities do not separate. See Figure~\ref{Uloop} for snapshots of NOAA 12715 during its evolution.  We speculate that this behavior is consistent with the emergence of an O-ring whose flux initially emerges but the polarities stay connected sub-surface. The categorization criteria we use show that 42$\%$ of the ARs have signatures consistent with formation via a kink instability or $\Sigma$-effect, 32$\%$ from multi-segment buoyancy, 16$\%$ from collisions and 11$\%$ unclassified but consistent with O-rings.

\section{Discussion \& Conclusions}

During the Solar Cycle 24 years of 2010-2019, 132 distinctly numbered sunspot groups were identified as containing a $\delta$-configuration as determined by forecasters and observers at NOAA and USAF. 18$\%$ of the $\delta$-spots were the simpler $\beta\delta$ category while the rest were the $\beta\gamma\delta$ configuration. As there were 1708 ARs numbered during Solar Cycle 24, and 1657 of them were observed by HMI/SDO, our values indicate that 8$\%$ of Cycle 24 spots were $\delta$-spots.  

Out of the 132 $\delta$-spots identified, we find that 46 produced flares of M-class strength and 14 produced flares of X-class strength as recorded in the GOES flare catalogue. While it is well known that $>$80$\%$ of X-class flares originate in $\delta$-regions \citep{guo:2014}, it is not commonly known that it is a small fraction of the $\delta$-spots that produce X-class flares, $\approx$10$\%$.  To restate: in Cycle 24, 8$\%$ of ARs are $\delta$-spots, and only $\approx$10$\%$ of those produce X-class flares, while 35$\%$ produce M-class flares. This is on the order of 10-20 ARs producing X-class flares and $\approx$50 producing M-class flares. 

Note that the $\delta$-spot sample has larger flux values than the control group.  Many super ARs (SARS) of every cycle are also $\delta$-spots and Cycle 24 is no different, as reported by \cite{chen:2016} who examines five SARS of Cycle 24 and all of those are also $\delta$-spots. Out of the largest 25 active regions of Cycle 24, 21 were $\delta$-spots, see  {https://www.spaceweatherlive.com/en/solar-activity/top-25-sunspot-regions/solar-cycle/24.html}. As such, the control group of $\beta$-spots has much smaller flux and this is a natural outcome of $\delta$-spots being large, in general.  Therefore, we could not curate a control group sample that mirrored the same flux range. 

\begin{figure}[!ht]
\begin{centering}
\includegraphics[trim=0.0in 0.0in 0.0in 0.0in,clip,width=0.70\textwidth]{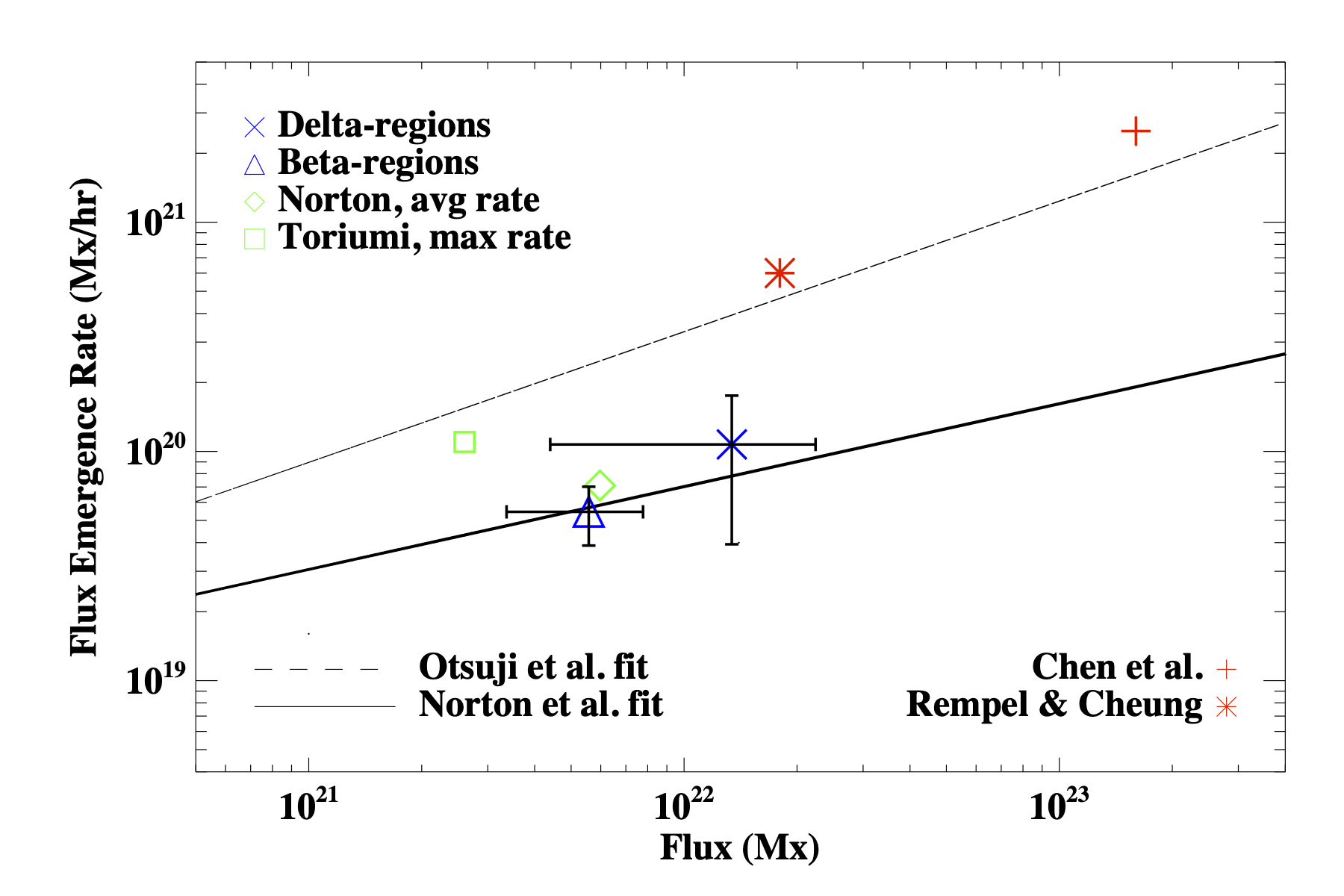}
\caption{AR flux emergence rates determined from observations (blue and green) and simulations (red) show the trend that regions containing more flux emerge faster.  The average values of AR maximum signed flux and average flux emergence rates in this study are the $\delta$-spots and $\beta$-spots shown as blue $\times$ and $\triangle$ symbols, respectively, with the median absolute deviation shown as error bars. 
We overlay fits using a dashed (solid) line reported by \cite{otsuji:PASJ} (\cite{Norton:2017}) of a power-law relationship from Hinode (HMI) observations where maximum (average) flux emergence
rates were dependent on maximum flux.  A value reported for the maximum flux emergence rate by \cite{Toriumi:2014} is plotted as a green $\square$ and an average flux emergence rate reported by \cite{Norton:2017} is a green $\triangle$. Two simulations of AR flux emergence from \cite{rempel:2014, chen:2017} are also shown using red symbols, see Sections 4 and 5 for more details and discussion.}
\end{centering}
\label{FE}
\end{figure}

\begin{figure}[!ht]
\begin{centering}
\includegraphics[trim=0.0in 0.52in 0.0in 0.4in,clip,width=0.74\textwidth]{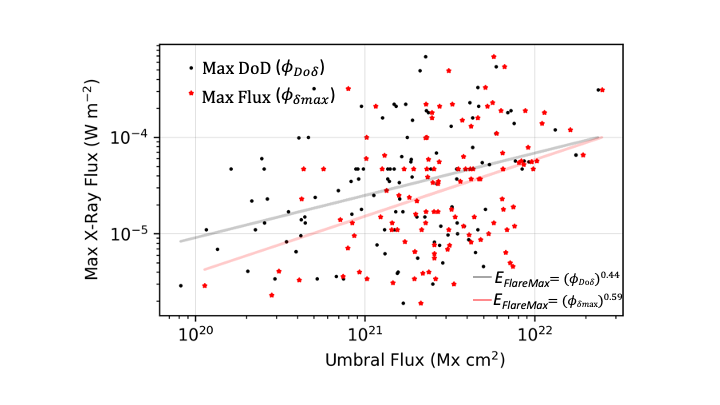}\\
\includegraphics[trim=0.0in 0.52in 0.0in 0.4in,clip,width=0.74\textwidth]{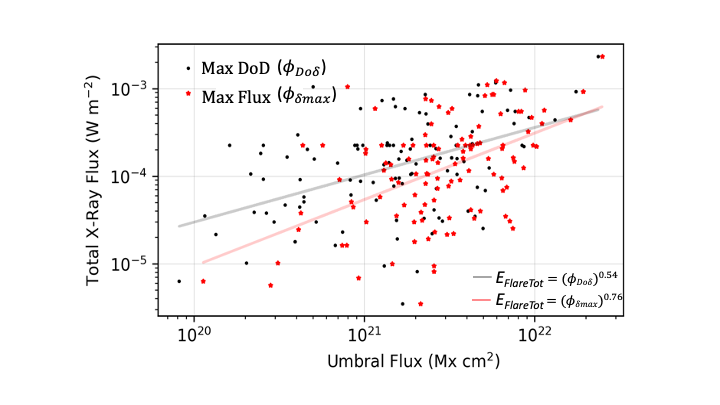}\\
\end{centering}
\caption{The maximum Do$\delta$ umbral flux, $\phi_{Do\delta}$ (black symbols), and the maximum total umbral flux, $\phi_{{\delta}_{max}}$ (red symbols), versus the single, maximum flare energy produced by that region is plotted at top for the large sample of $\delta$-spots.  The same quantities, $\phi_{Do\delta}$ and $\phi_{{\delta}_{max}}$, are plotted versus total observed flare energy over time at bottom. 
The best fits are given in the legend and discussed in Section 4.  All fits provide a positive correlation but the correlation between maximum umbral flux, $\phi_{{\delta}_{max}}$ and total X-ray flux, $E_{FlareTot}$ has the highest r-value of 0.52.
} 
\label{flares}
\end{figure}

The tilt angles reported in Tables 1-2 and Figure~\ref{anti-hale} indicate that $\delta$-spot formation processes could be responsible for a significant fraction of all AH regions observed during a solar cycle.  The total percentage of ARs in any given cycle that are AH is roughly 4$\%$, as reported by \citet{wang:1984, kosovichev:2008}, or 8$\%$, as reported by \citet{mcclintock:2014, li:2018}.  We find that 44$\%$ of the small sample of $\delta$-spots are AH when using the $\delta$-umbrae only to determine tilt and 16$\%$ of $\delta$-regions are anti-Hale when using all umbrae to determine the tilt. If we use these values of 8$\%$ of ARs in Solar Cycle 24 were $\delta$-spots, and 16$\%$ of those were AH, then the AH $\delta$-spots account for 1.3$\%$ of all ARs. Comparing this with 4-8$\%$ of ARs in Solar Cycle 24 being AH indicates that $\delta$-spots account for $\approx$15-30$\%$ of all AH regions during Solar Cycle 24.  We find that our percentage of AJ and AH  (when considering all umbrae to determine tilt as in Table 2 as this is the correct comparison for the references above) is 53$\%$ in the small sample and 50$\%$ in the large sample, which is higher than the 34$\%$ reported by \cite{Tian:2005}. Note that the tilt angles reported in this paper are determined at only the times of  $\phi_{Do\delta}$ and $\phi_{{\delta}_{max}}$. The percentages of $\delta$-spots with AH or AJ tilts may certainly be higher if we considered the tilt values at all times during the $\delta$-spot lifetime. 

The ARs spend, on average, just over half of their observed time in the $\delta$ configuration and were seen to exist for only a quarter of the lifetime in a few regions. Since the $\delta$-configuration can exist for a short time (even less than a day) compared to the total AR lifetime, we suggest using an automated analysis code with space-based data as input in order to capture more instances of ARs in $\delta$-configuration.  This would also remove operator error when classification is performed by humans or ambiguity introduced by seeing conditions at ground-based observatories.  

It is not unexpected to find a higher rotation rate in the $\delta$-spot sample as this has been reported previously.  The average AR group (not $\delta$) experiences a tilt angle change of $\approx$5$\degree$ per day and has a 40$-$50 Mm footpoint separation \citep{mcclintock:2016}  within a day of emergence.  The values from Table 1 indicate that the $\delta$-portion of the spot averages a 62$\degree$ rotation over 88 hours, or 17$\degree$ per day.  The $\beta$ spot values from Table 3 show a rotation of 9$\degree$ over 99 hours which is a rotation on the order of 2$\degree$ per day. Therefore, we can say that the $\delta$-only portion, the magnetic knot, of the umbrae rotates at a rate 8-9$\times$ higher than other categories of active regions. However, this result should be confirmed by more sophisticated analysis using careful feature tracking rather than the simple change in tilt angle as we have done.

Another interesting finding is that the the polarity separations of the $\beta$-spots and the $\delta$-spots (when considering all umbrae as in Table 2) are similar, i.e. 66$\pm$16 Mm for $\beta$s and 61$\pm$16 Mm for $\delta$s. $\delta$-spots are often described in the literature as ``compact" but this is a term best used to describe the ``island"-$\delta$s.  In fact \citet{Zirin:1987} only uses the term as applied to those types, see Figure 1a, and not the quadrupolar regions. In addition, the large flux of $\delta$-spots means that even if a region is compact, the separation distance of the polarity centroids can be sizeable due to the large area of the region.  
 
 Because flux emergence rates are dependent on the amount of flux in the system, the $\delta$ flux emergence rate is higher, as expected, than for the $\beta$-spots because the $\delta$s have 2.6$\times$ as much flux as the $\beta$s. The average $\delta$ flux emergence rate can still be considered within the normal range for the maximum flux associated with the active regions since the emergence rate scales as $\dot{\phi} \propto \phi_{max}^{0.36}$ \citep{Norton:2017}. This is shown in Figure 8 - note that the $\beta$ and $\delta$ emergence rates lie directly on the line. 
 
 Simulations of AR flux emergence have been carried out by numerous researchers, see the review by \cite{2014LRSP...11....3C}, and for results that easily compare to observations, see the introduction section of \cite{Norton:2017}. In general, simulations show both a faster flux emergence and a shorter flux emergence time period than observations, see \cite{rempel:2014}, \cite{chen:2021arXiv210614055C, Knizhnik:2022a}.  Results from simulations are shown in Figure 8 (red symbols) and are compared to the observational results from this paper (blue symbols) and previous observational results (green symbols) \citep{Toriumi:2014, Norton:2017}.  Note that changing the noise limit for inclusion of more flux in the observations, and calculating the maximum instantaneous flux emergence rate as reported by \cite{otsuji:PASJ} (dashed line in Figure 8 showing the relationship of $d{\phi}/dt \propto \phi^{0.57}$ based on Hinode SOT observations of events followed in time for a few hours), appears to close the gap between the observational flux emergence rate and those reported in simulations.  

Interestingly, \cite{chen:2017, chen:2021arXiv210614055C} found that the depth at which the flux was introduced into their convective simulations affected the emergence rate with flux introduced in deeper domains emerging slower and generally matching the convective upflow speeds at that depth. \cite{chen:2017} found that less flux emerged in the photosphere when it was initially placed deeper, as well.  ARs are almost certainly emerging from flux that is generated deeper than 10 or 20 Mm below the photosphere, which is the extent of many simulations whose depths are limited by available computing power for the simulations. However, it is not definitely known at what depth the magnetic field is amplified in the dynamo process.  It seems highly coincidental that the size scale of active regions, with polarity footpoint separations of $\approx$40 Mm in the first few days of emergence and separations of $\approx$70 Mm in its last days of decay \citep{mcclintock:2016}, are the same size of a supergranule or two \citep{Rincon:2018}. The depth of the near surface shear layer, ~35 Mm, is also the horizontal extent of a supergranule \citep{Matilsky:2018}.  We mention these depths and extents in order to encourage modelers to consider the interactions between supergranules and flux tubes in the 20 - 100 Mm below the surface.

The difficulties of simulating flux rising throughout the bulk of the convection zone and coupling it to the upper-most 50 Mm below the photosphere are due to the change from anelastic approximations to fully compressible realizations as well as the rapid change of pressure scale heights which is many orders of magnitude.  Therefore, the full story of magnetic field amplification and emergence is not known and the near-surface shear layer's role remains a mystery.

\begin{figure}[hpt]
\hbox to 5cm{\hfil Sign of current helicity as a function of time and latitude \hfil}\par
\includegraphics[trim=0.0in 0.10in 0.0in 0.1in,clip,width=0.7\textwidth]{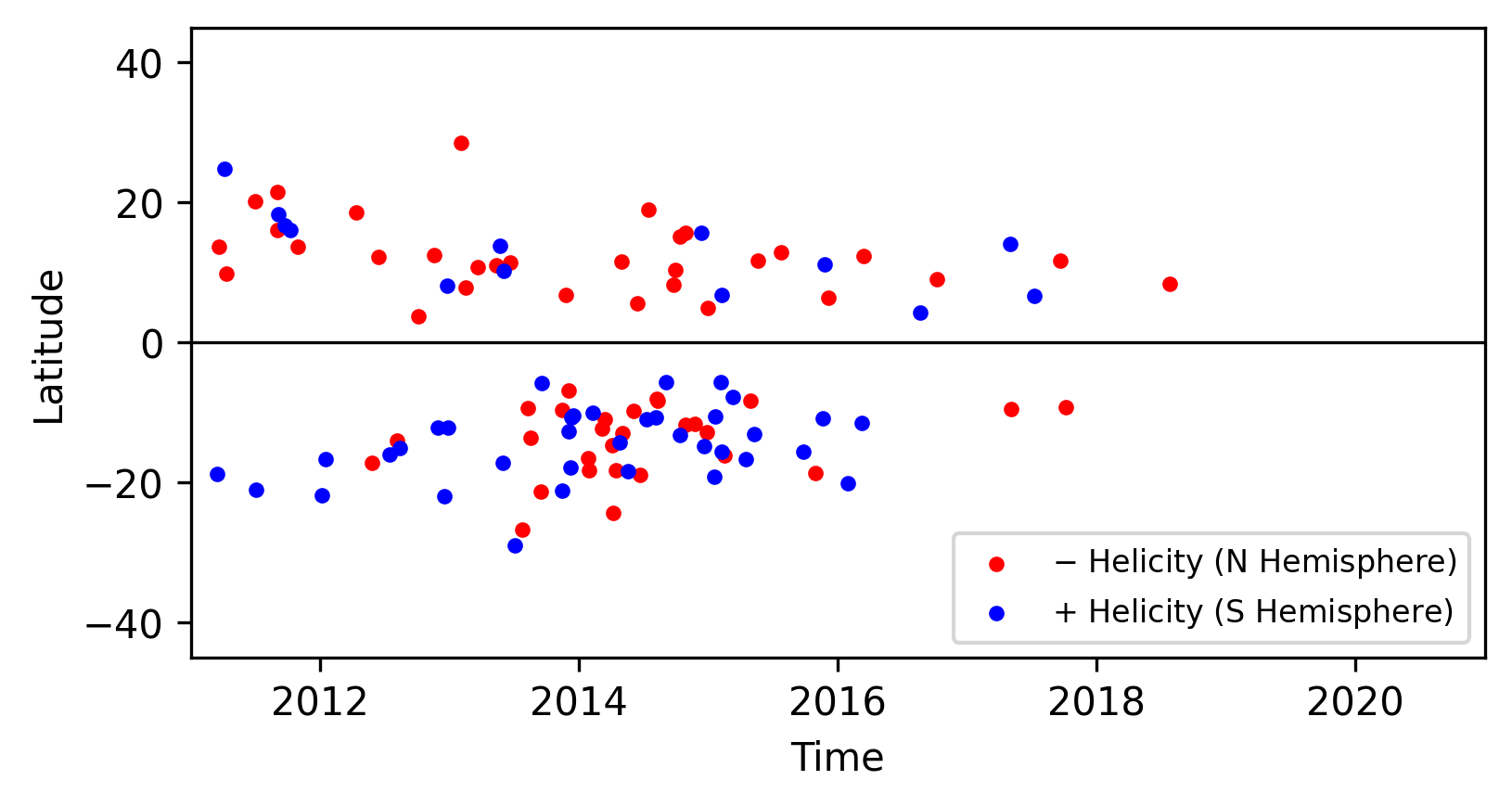}\\
\includegraphics[trim=0.0in 0.1in 0.0in 0.1in,clip,width=0.7\textwidth]{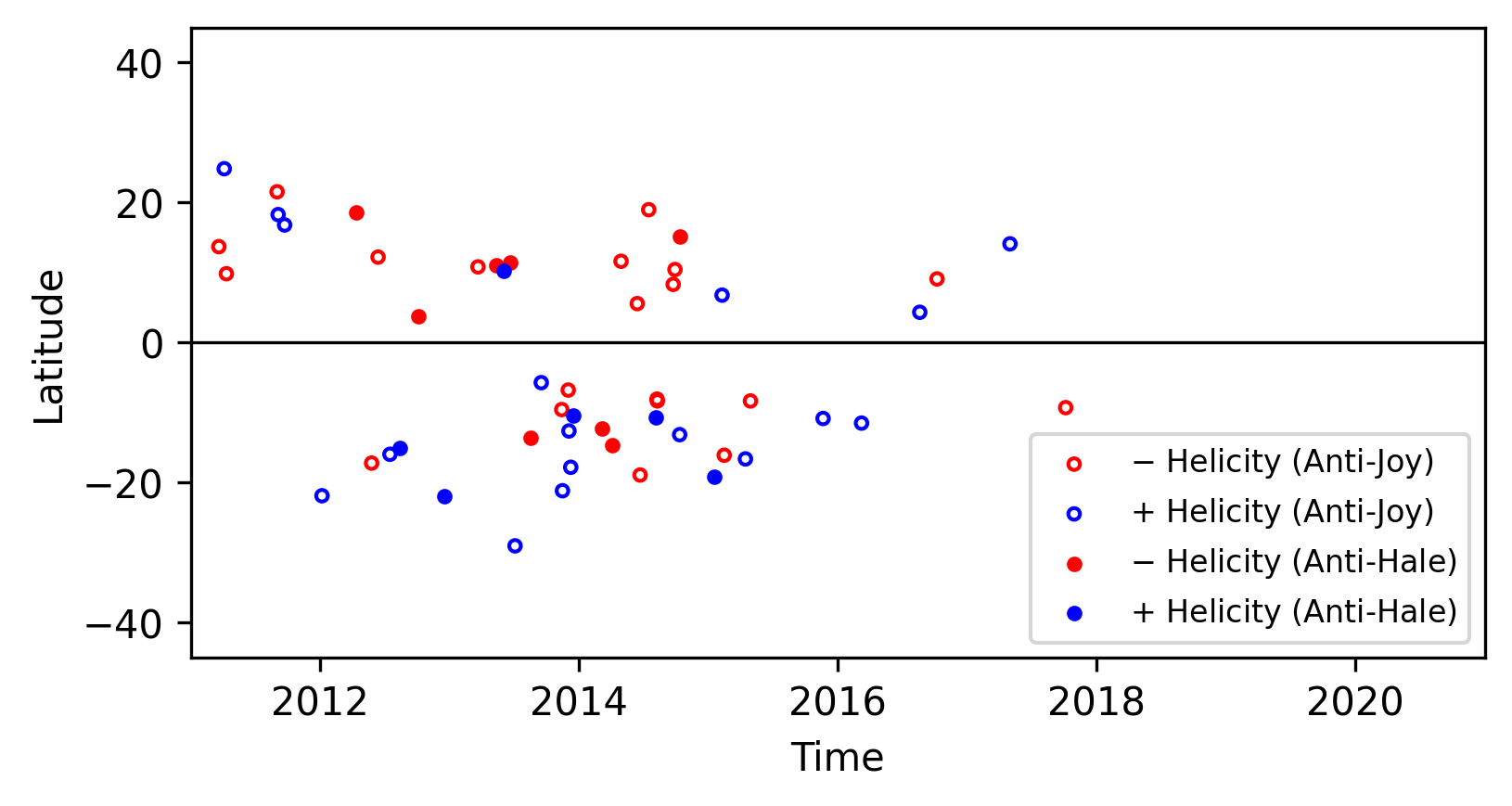}\\
\caption{The sign of current helicity for the $\delta$-regions is shown as a function of time and latitude.  64$\%$ of all the $\delta$-regions follow the hemispheric rule (HR) in which the Northern (Southern) hemispheric ARs have a preference for $-$ ($+$) current helicity. The Northern hemisphere has 71$\%$ obeying the HR whereas the Southern hemisphere has 57$\%$. We found that 74$\%$ of anti-Hale $\delta$-regions exhibit the preference (83$\%$ in N and 63$\%$ in S) as do 60$\%$ of the anti-Joy regions (65$\%$ in N and 55$\%$ in S).    }
\label{helicity}
\end{figure}

\begin{table}[!ht]
	\begin{center}
	    \begin{tabular}{|c|c|c|c|c|c|c||c|c|c|c|}
			\hline
			NOAA (HARP) & Time & AH & AJ & Do$\delta$ & $\phi_{Do\delta}$ & Sep& $\Delta$Rot & $\Delta$Sep & Life & $\Sigma$Flares \\
			\hline
			& & \multicolumn{5}{c||}{Instantaneous Values at Max Do$\delta$}&\multicolumn{4}{c|}{Time-derived Values}\\
			\hline
			11158 (377) & 2011.02.15 14:00 & 0 & 0 & 54 & 2.25 & 32.60 & 53 & 0.10 & 80 & 5.90 \\
			\hline
			11166 (401) & 2011.03.10 13:36 & 0 & 1 & 41 & 2.00 & 81.10 & 57 & 2.70 & 40 & 2.60 \\
			\hline
			11261 (750) & 2011.08.02 17:24 & 1 & 0 & 77 & 2.27 & 37.20 & 120 & -1.80 & 68 & 3.60 \\
			\hline
			11263 (753) & 2011.08.02 10:36 & 1 & 0 & 40 & 3.13 & 13.20 & 55* & -3.30 & 76 & 8.60 \\
			\hline
			11267 (764) & 2011.08.07 08:12 & 1 & 0 & 86 & 0.28 & 5.70 & -40 & 0.20 & 12 & 0.10 \\
			\hline
			11302 (892) & 2011.09.29 18:48 & 0 & 1 & 55 & 4.81 & 15.20 & -85 & -1.90 & 176 & 9.70 \\
			\hline
			11429 (1449) & 2012.03.06 00:24 & 1 & 0 & 99 & 10.00 & 35.80 & 35 & 1.80 & 140 & 12.00 \\
			\hline
			11465 (1596) & 2012.04.23 04:00 & 0 & 1 & 76 & 1.95 & 20.40 & -35 & 3.70 & 32 & 0.31 \\
			\hline
			11476 (1638) & 2012.05.10 02:12 & 0 & 1 & 89 & 10.30 & 28.50 & 100* & 23.80 & 130 & 5.50 \\
			\hline
			11520 (1834) & 2012.07.09 20:24 & 0 & 1 & 78 & 11.20 & 52.70 & 5 & -8.20 & 124 & 4.00 \\
			\hline
			11560 (1993) & 2012.09.02 23:12 & 1 & 0 & 58 & 1.11 & 9.60 & 77* & -1.10 & 42 & 0.61 \\
			\hline
			11598 (2137) & 2012.10.25 20:00 & 0 & 1 & 99 & 3.30 & 30.50 & $-$ & 13.60 & 78 & 4.00 \\
			\hline
			12158 (4536) & 2014.09.09 03:00 & 1 & 0 & 99 & 6.47 & 38.60 & 10 & 5.00 & 46 & 2.30 \\
			\hline
			12192 (4698) & 2014.10.26 17:36 & 0 & 0 & 95 & 32.90 & 93.00 & 30 & -25.60 & 200 & 23.00 \\
			\hline
			12205 (4781) & 2014.11.07 05:24 & 1 & 0 & 65 & 2.33 & 3.30 & 88* & -4.20 & 134 & 7.30 \\
			\hline
			12443 (6063) & 2015.10.31 14:12 & 0 & 0 & 38 & 2.95 & 41.50 & $-$ & -12.40 & 58 & 2.40 \\
			\hline
			12671 (7107) & 2017.08.18 03:12 & 1 & 0 & 30 & 1.43 & 14.30 & -180 & 3.70 & 102 & 0.71 \\
			\hline
			12673 (7115) & 2017.09.06 10:12 & 0 & 1 & 93 & 13.00 & 21.70 & $-$ & 10.70 & 84 & 30.00 \\
			\hline
			12715 (7275) & 2018.06.21 13:00 & 0 & 0 & 93 & 1.17 & 34.50 & 20 & -19.10 & 52 & 0.06 \\
			\hline
			Average & $-$ & 42$\%$ & 36$\%$ & 72 & 5.94 & 32.10 & 62 & -0.66 & 88 & 6.46 \\
			\hline
			Median & $-$ & $-$ & $-$ & 77 & 2.95 & 30.50 & 54 & 0.07 & 78 & 4.00 \\
			\hline
			Med Abs Dev & $-$ & $-$ & $-$ & 19 & 1.78 & 11.00 & 43 & 3.59 & 36 & 3.30 \\
			\hline
		\end{tabular}
	\end{center}
	\caption{Parameters of $\delta$-spots at maximum umbral flux participating in the $\delta$-configuration,  $\phi_{Do\delta}$, and some time derived values. Columns 1-2 correspond to the NOAA and HARP number.  The columns 3-7 correspond to instantaneous values at the time of maximum Do$\delta$ for this region: if the bipole was anti-Hale or anti-Joy (1 yes, 0 no) at that time, the Do$\delta$ in $\%$, the umbral flux participating in the $\delta$ ($\phi_{Do\delta}$ given in units of 1$\times$10$^{21}$ Mx), and the separation (in Mm). Columns 8-11 contain time-derived values. Column 8 contains the rotation for individual knots which is positive (negative) if it is counter-clockwise (clockwise). An $\ast$ indicates an average of several knots rotating in both directions, and $-$ indicates there was no measurable rotation. The average value of the rotation rates are calculated from the absolute values. Column 9 shows separation in Mm. The  separation is calculated from the first to the last time the $\delta$-configuration exists. Column 10 shows the lifetime (in hours) and is how long the $\delta$-configuration exists. Column 11 is the sum of flare energy from this AR. The last row is the absolute deviation of the median.}
\end{table}

We had hoped that the Do$\delta$ and $\phi_{Do\delta}$, would be positively correlated with single maximum or total flare energy of that active region.  While the correlation was positive, the maximum flux of all umbrae, $\phi_{{\delta}_{max}}$, provided a slightly higher r-value (0.52) in the fit to total flare energy than the fit using $\phi_{Do\delta}$ (r=0.47), see Figure~\ref{flares}.  This was disappointing but it has been pointed out previously by \cite{fisher:1998} that X-ray luminosity of a non-flaring AR is best correlated with total unsigned magnetic flux of an AR as opposed to other magnetic quantities.  \cite{sammis:2000} showed that both the spot group area and the increasing magnetic complexity of an AR was positively correlated with the single highest maximum flare energy associated with that region. The Figure~\ref{flares} plots are a representation of the upper, right portion of the plot by \cite{sammis:2000} which showed maximum flare energy for all classifications of sunspot, but our plot shows only the Cycle 24 $\delta$s. If we isolated the source region of the flare to specific umbrae, it is possible that the flux participating in the $\delta$ and the Do$\delta$ would be more flare-rich than the umbrae not participating in the $\delta$, but this is left to a future study.

Our findings that the $\delta$-spots obey the hemispheric rule (HR) for current helicity is in agreement with many other studies on this topic, ie., that ARs obey the HR roughly one-half to three-quarters of the time \citep{pevtsov:1995}.  Our results also agree that the Northern hemisphere adheres to the rule slightly more often than the Southern hemisphere in the last several cycles.  We speculated that the anti-Hale or anti-Joy regions might not obey the HR as often, but the lower plot in Figure~10 debunks this hypothesis. As the current helicity measure is not determined for the $\delta$-only umbrae, we may find a higher deviation from the HR if we were able to calculate the current helicity for the $\delta$-umbrae distinct from the entire active region. 

\begin{table}[!ht]
	\begin{center}
		\begin{tabular}{|c|c|c|c|c|c|c||c|c|c|c|c|}
			\hline
			NOAA (HARP) & Time & AH & AJ & DoD & $\phi_{{\delta}_{max}}$ & Sep & $\dot{\phi}_{{\delta}_{max}}$ & $\Delta$Rot & $\Delta$Sep & Life & $\Sigma$Flares \\
			\hline
			& & \multicolumn{5}{c||}{Instantaneous Values at Max Do$\delta$}&\multicolumn{5}{c|}{Time-derived Values}\\
			\hline
			11158 (377) & 2011.02.15 06:00 & 0 & 0 & 33 & 19.50 & 66.60 & 7.35 & 20 & 33.10 & 160 & 5.90 \\
			\hline
			11166 (401) & 2011.03.10 17:36 & 0 & 1 & 35 & 25.20 & 99.10 & 4.74 & 5 & 9.90 & 198 & 2.60 \\
			\hline
			11261 (750) & 2011.08.02 17:24 & 1 & 0 & 77 & 13.60 & 19.70 & 0.00 & 0 & -39.80 & 164 & 3.60 \\
			\hline
			11263 (753) & 2011.08.01 20:36 & 0 & 1 & 0 & 20.80 & 65.90 & 11.10 & -6 & 24.30 & 206 & 8.60 \\
			\hline
			11267 (764) & 2011.08.05 22:12 & 0 & 0 & 0 & 3.84 & 31.00 & 3.78 & -5 & 38.50 & 60 & 0.10 \\
			\hline
			11302 (892) & 2011.09.24 23:24 & 0 & 0 & 21 & 41.60 & 93.90 & 0.00 & 10 & -77.10 & 180 & 9.70 \\
			\hline
			11429 (1449) & 2012.03.06 04:24 & 1 & 0 & 89 & 36.70 & 39.00 & 3.75 & 40 & 87.60 & 190 & 12.00 \\
			\hline
			11465 (1596) & 2012.04.21 18:00 & 0 & 1 & 0 & 9.32 & 54.90 & 6.55 & -35 & -1.60 & 188 & 0.31 \\
			\hline
			11476 (1638) & 2012.05.09 21:36 & 0 & 1 & 73 & 37.60 & 48.90 & 0.00 & 10 & -119.90 & 188 & 5.50 \\
			\hline
			11520 (1834) & 2012.07.09 08:24 & 0 & 1 & 77 & 61.60 & 62.50 & 0.00 & 5 & 83.90 & 184 & 4.00 \\
			\hline
			11560 (1993) & 2012.09.01 17:12 & 0 & 1 & 0 & 9.11 & 70.10 & 3.66 & -50 & 27.10 & 144 & 0.61 \\
			\hline
			11598 (2137) & 2012.10.23 05:00 & 0 & 0 & 28 & 21.20 & 45.70 & 0.00 & -5 & -14.50 & 120 & 4.00 \\
			\hline
			12158 (4536) & 2014.09.09 05:00 & 1 & 0 & 95 & 21.40 & 38.60 & 37.30 & 10 & -12.90 & 120 & 2.30 \\
			\hline
			12192 (4698) & 2014.10.25 06:48 & 0 & 0 & 76 & 112.00 & 96.50 & 12.70 & -15 & 19.80 & 200 & 23.00 \\
			\hline
			12205 (4781) & 2014.11.07 05:24 & 0 & 0 & 65 & 23.80 & 40.50 & 0.00 & 30 & 55.40 & 170 & 7.30 \\
			\hline
			12443 (6063) & 2015.10.31 18:12 & 0 & 0 & 9 & 37.20 & 116.00 & 0.00 & -10 & -67.80 & 172 & 2.40 \\
			\hline
			12671 (7107) & 2017.08.17 21:12 & 0 & 0 & 28 & 18.40 & 100.00 & 4.51 & -5 & 143.50 & 170 & 0.71 \\
			\hline
			12673 (7115) & 2017.09.06 20:36 & 0 & 1 & 75 & 39.80 & 27.50 & 22.30 & -150 & -9.50 & 110 & 30.00 \\
			\hline
			12715 (7275) & 2018.06.20 08:48 & 0 & 0 & 0 & 5.67 & 44.10 & 11.00 & 20 & -5.80 & 144 & 0.06 \\
			\hline
			Average & $-$ & 15$\%$ & 36$\%$ & 41 & 29.40 & 61.10 & 10.41 & 23 & 9.18 & 160 & 6.46 \\
			\hline
			Median & $-$ & $-$ & $-$ & 33 & 21.40 & 54.90 & 6.55 & 10 & 9.95 & 170 & 4.00 \\
			\hline
			Med Abs Dev & $-$ & $-$ & $-$ & 33 & 12.30 & 15.90 & 2.8 & 5 & 24.40 & 20 & 3.30 \\
			\hline
		\end{tabular}
	\end{center}
	\caption{Parameters of $\delta$-spots shown at the time of maximum umbral flux ($\phi_{{\delta}_{max}})$ of all umbrae (not only of the $\delta$-portion) and some time averaged values.  Columns 1-2 correspond to the NOAA and HARP number.  Columns 3-7 correspond to instantaneous values of the time of maximum flux contained in all the umbrae: if the bipole (as defined by the centroids determined using all umbrae) was anti-Hale or anti-Joy (1 yes, 0 no) at that time, the Do$\delta$ in $\%$, the unsigned umbral flux participating in the entire AR ($\phi_{{\delta}_{max}}$) given in units of 1$\times$10$^{21}$ Mx, and the separation (in Mm). Columns 8-12 contain time-derived values. Column 8 shows the flux emergence rate which is the total flux that emerged from a time nearest 10$\%$ and through to 90$\%$ of maximum umbral flux divided by the number of hours in which that occurred, then divided by two to estimate the signed flux in the emerging flux tube. Column 9 contains the rotation rate for the entire AR which is positive (negative) if it is counter-clockwise (clockwise). Column 10 shows separation in Mm. The  separation is calculated from the first to the last time the AR is observed.  Column 11 is the lifetime (in hours) of how long the umbrae exist. Lastly, in column 12, the sum of flare energy from this AR is shown. Note 11560 is anti-Hale when considering only umbrae in $\delta$-configuration (see Table 1) but anti-Joy when considering all umbrae (as shown here). The last row is the absolute deviation of the median. }
\end{table}

We find that active regions classified as $\delta$ often contain a significant fraction of umbrae that are not participating in the $\delta$.  On average, the  regions have a maximum Do$\delta$ of 72$\%$ and spend only 55$\%$ of their time on the disk as a $\delta$.  The calculated tilt angles of the $\delta$-portion are found in 37$\%$ of the regions to be in completely different quadrants than the tilt angles calculated using all umbrae of the region. As an example, see the fourth panel from the top in Figure 2, in which the $\delta$-configuration is comprised of the opposite polarities with an anti-Hale tilt in the middle of an extended quadrapolar region presenting an anti-Joy tilt.  These geometries could arise due to the fact that a kink instability is acting on only a portion of the flux tube or there is interaction between the following and leading polarities of a multi-segment buoyancy.  

In our first categorization, we quantify how many regions are single or multiple emergence events and how many are bipoles or quadrupoles, see Table 4.  It is surprising to find that 84$\%$ of the $\delta$-spots are formed in a single flux emergence event and over half, 58$\%$, of $\delta$-regions are formed as quadrupoles, SEEQ.  There are fewer bipoles, 26$\%$, and fewer collisions, 16$\%$, than we expected.  

Some of these ARs, such as 11302 and 12443, do not display AH or AJ. 12158 is compact with an AH tilt with a high Do$\delta$ so it appears consistent with a kink instability but has very little rotation. 12715 is unusual because it is not AH nor AJ, shows very little rotation, emerges with signatures similar to an arch but the polarities do not separate.  We speculate that this behavior is caused by a loop connected underneath the surface. 

Our second categorizing of the $\delta$-regions into probable formation types is more speculative since we cannot with confidence distinguish between several formation types using our current measures.  The categories are kink instability or $\Sigma$-effect, multi-segment buoyancy (labeled as ``quadrupole" or ``spot-satellite" in Table 4), or interacting/colliding ARs using observed characteristics reported herein.  Our percentages of categorization roughly agree with \cite{ToriumiMagProp:2017}, with our percentages being 42$\%$ kink instability or $\Sigma$-effect, 32$\%$ multi-segment buoyancy and 16$\%$ collision, while \cite{ToriumiMagProp:2017} found 39$\%$ kink-instability, 55$\%$ multi-segment buoyancy and 6$\%$ collision.  \cite{ToriumiMagProp:2017} used measures such as the location and length of the polarity inversion line, flare ribbons and proper motion for categorization.  Our sample contained eight of the same sunspots (out of 31 in the \cite{ToriumiMagProp:2017} study and 19 herein) and we only categorized half of those in the same manner. We categorized ARs 11158 and 11476 as likely formed by a kink-instability when \cite{ToriumiMagProp:2017} labeled them as multi-segment buoyancy and vice versa for ARs 11429 and 12192.  Interestingly, ARs 11465 and 12715 behave in a manner that is consistent with a $\delta$-spot being formed by a rising, sub-surface O-ring, with no anti-Hale tilt but the bipole emerges as an arch similar to other rising flux loops. However, once that arch has emerged, there is no separation of the polarity footpoints over time and no real rotation, as if the regions are still connected sub-surface via a flux tube with very little writhe (see \citet{spruit:1987} for a sketch of a 'repairing' active region loop whose legs reconnect to form an O-ring). This raises the question as to how commonly this type of $\delta$-spot is seen. 

\begin{table}[t]
	\begin{center}
		\begin{tabular}{|c|c|c|c|c|c|c||c|c|c|c|c|}
			\hline
			NOAA (HARP) & Time & AH & AJ & Do$\delta$ & $\phi_{{\beta}_{max}}$ & Sep & $\dot{\phi}_{\beta}$ & $\Delta$Rot & $\Delta$Sep & Life & $\Sigma$Flares \\
			\hline
			& & \multicolumn{5}{c||}{Instantaneous Values at $\phi_{{\beta}_{max}}$}&\multicolumn{5}{c|}{Time-derived Values}\\
			\hline
			11141 (325) & 2010.12.31 15:36 & 0 & 0 & 0 & 5.31 & 44.70 & 0.92 & -6 & 29.40 & 108 & 0.06 \\
			\hline
			11184 (466) & 2011.04.05 21:00 & 0 & 0 & 0 & 21.60 & 97.50 & 7.42 & -17 & 26.90 & 86 & 0.02 \\
			\hline
			11199 (540) & 2011.04.28 18:00 & 0 & 0 & 0 & 11.90 & 85.10 & 6.54 & -2 & 51.90 & 70 & 0.18 \\
			\hline
			11311 (926) & 2011.10.04 15:24 & 0 & 0 & 0 & 4.30 & 57.90 & 8.27 & 0 & 9.50 & 20 & 0.04 \\
			\hline
			11327 (982) & 2011.10.21 18:00 & 0 & 0 & 0 & 8.36 & 54.50 & 4.51 & 2 & 45.50 & 100 & 0.04 \\
			\hline
			11397 (1312) & 2012.01.13 08:12 & 0 & 1 & 0 & 4.23 & 25.70 & 7.68 & 20 & 25.50 & 22 & 0.04 \\
			\hline
			11428 (1447) & 2012.03.06 14:00 & 0 & 0 & 0 & 10.10 & 60.60 & 4.58 & 5 & 10.50 & 138 & 0.16 \\
			\hline
			11435 (1471) & 2012.03.17 22:36 & 0 & 0 & 0 & 10.30 & 65.90 & 4.94 & 10 & 39.90 & 64 & 0.16 \\
			\hline
			11460 (1578) & 2012.04.21 14:36 & 0 & 0 & 0 & 16.20 & 87.00 & 5.97 & 11 & 56.20 & 154 & 0.12 \\
			\hline
			11512 (1795) & 2012.06.30 21:48 & 0 & 0 & 0 & 11.20 & 87.70 & 5.23 & -14 & 38.10 & 154 & 0.14 \\
			\hline
			11497 (1727) & 2012.06.05 09:12 & 0 & 0 & 0 & 19.10 & 60.40 & 3.91 & 13 & 14.90 & 172 & 0.04 \\
			\hline
			Average & $-$ & 0 & 9$\%$ & 0 & 11.10 & 66.10 & 5.45 & 9 & 31.70 & 99 & 0.09 \\
			\hline
			Median & $-$ & $-$ & $-$ & 0 & 10.30 & 60.60 & 7.87 & 10 & 29.42 & 100 & 0.06 \\
			\hline
			Med Abs Dev & $-$ & $-$ & $-$ & 0 & 4.99 & 15.90 & 6.14 & 5 & 14.60 & 38 & 0.04 \\
			\hline
		\end{tabular}
	\end{center}
\caption{Parameters of $\beta$-spots at maximum umbral flux, $\phi_{{\beta}_{max}}$ with the parameters as described in Table 2 but for this table using the flux in the $\beta$-spot umbrae.}
\end{table}
The difference between an inverted kink instability (upper right, Figure 1) and a multi-segment buoyancy configuration such as a quadrupole (lower right, Figure 1) is uncertain.  While numerical simulations of the kink instability do not produce structures with inverted kinks, observations of emerging flux regions are highly suggestive of such a configuration. \citet{Takizawa:2015} identify ``downward knotted structures in the middle part of the magnetic flux tube" in a dozen $\delta$-spots.

In order to more confidently identify that a kink instability is responsible for the formation of a $\delta$-configuration, one should examine the twist (either using the current helicity measure or another parameter) and writhe.  The kink instability is implicated as the formation mechanism when the twist and the writhe have the same sign.  This is in contrast to the $\delta$-configuration being formed during magnetic flux tube interactions with turbulent convection ($\Sigma$-effect) in which case the twist and writhe would have opposite signs.  Previous research has concluded that $\delta$-spots are formed by multiple mechanisms since only a portion of the studied regions have the same sign of twist and writhe. \citet{fuentes:2011} found 6 out of 10 island-$\delta$ regions had the same sign of twist and writhe and therefore were consistent with a kink instability while the remaining 4 regions had opposite signs of twist and writhe. \citet{Tian:2005} reported a similar result with $\approx$65$\%$ of 107 $\delta$-spots having similar signs of twist and writhe.

\citet{Knizhnik:2018} simulated the emergence of kink unstable flux ropes and found that while the writhe was identifiable from surface measurements in the simulations, and was consistent with the kink formed in the convection zone, the twist parameter alpha was not coherent and did not give a clear signature either consistent or inconsistent with the kink. The conclusion they drew was that the dramatic dynamics of emergence and expansion into the corona distort the field enough that the twist parameter signature is not representative of the flux rope's twist.

A more in-depth study on the twist, writhe and current helicity of the magnetic knots of $\delta$-spots is warranted. However, it is unwise to measure the twist and writhe of the entire AR when only small portions of the AR are participating in the $\delta$-configuration.  Out of the 19 $\delta$-spots in our small sample, only 2 of them (not including the O-ring regions) could be considered to have the majority of the AR be participating in the magnetic knot. Meaning, these can be considered an island $\delta$ type configuration in which the AR rotates bodily instead of having several small knots within the region with distinctly different rotations and dynamics.  Those two regions are NOAA 12158 and 11520 with magnetic knot separations of 30-50 Mm and low total rotations of 5-10$\degree$ in which the total AR rotates the same amount as the magnetic knot.  Most of the magnetic knots are smaller than the entire AR.  For example, NOAA 11267, 11560 and 12671 have knot separations on the order of 5-15 Mm with rotations ranging from 40 - 180$\degree$.  These knots should be studied individually and distinct from the behavior of the AR in which they are embedded as the kink instability may be acting on small flux tubes within the AR.  We hope to do this in a subsequent publication as it is beyond the scope of this paper.


\begin{table}
	\begin{center}
		\begin{tabular}{|c|c|c|c|c|c|c||c|c|}
			\hline
			NOAA (HARP) & AH & AJ & Do$\delta$  & ${\Delta}$Rot & Quad. & Em. Events & Categorization 1 & Categorization 2 \\
			\hline
			11158 (377) & 0 & 0 & 61  & 53  &1& 1 & SEEQ& Inverted Kink (Quadrupole) \\
			\hline
			11166 (401) & 0 & 1 & 50  & 57 &1 & $>$1 & MEEQ & Colliding  \\
			\hline
			11261 (750)  & 1 & 0 & 78 & 120 & 1& 1 & SEEQ& Quadrupole (Inverted Kink) \\
			\hline
			11263 (753) & 1 & 0 & 40 & 55*  &1& $>$1 & MEEQ & Colliding \\
			\hline
			11267 (764) & 1 & 0 & 77  & -40  & 1&1 & SEEQ& Inverted Kink (Quadrupole)\\
			\hline
			11302 (892) & 0 & 0 & 83  & -85 &1& 1 & SEEQ&  Quadrupole \\
			\hline
			11429 (1449) & 1 & 0 & 100  & 35 &1& 1& SEEQ& Quadrupole \\
			\hline
			11465 (1596) & 0 & 1 & 76 & -35 &0 & 1 & SEEB& O-ring\\
			\hline
			11476 (1638) & 0 & 1 & 93 & 100* & 1& 1 & SEEQ & Multiple Kink \\
			\hline
			11520 (1834) & 0 & 1 & 79 & 5 &  0&1& SEEB& Kink or $\Sigma$-effect\\
			\hline
			11560 (1993) & 1 & 0 & 56 & 77* &1 & 1 & SEEQ& Quadrupole (Inverted Kink)\\
			\hline
			11598 (2137) & 0 & 1 & 99  & $-$ & 0&1 & SEEB& Kink or $\Sigma$-effect\\
			\hline
			12158 (4536) & 1 & 0 & 100  & 10 &0& 1 & SEEB& Kink or $\Sigma$-effect\\
			\hline
			12192 (4698) & 0 & 0 & 95 & 30 & 1&1& SEEQ& Spot Satellite \\
			\hline
			12205 (4781) & 1 & 0 & 65  & 88* & 1 & 1& SEEQ& Inverted Kink (Quadrupole) \\
			\hline
			12443 (6063) & 0 & 0 & 39 & $-$ & 1& 1 & SEEQ&  Quadrupole \\
			\hline
			12671 (7107) & 1 & 0 & 30 & -180 & 1 & 1& SEEQ& Inverted Kink (Quadrupole)  \\
			\hline
			12673 (7115)  & 0 & 1 & 94 & $-$ &1& $>$1 & MEEQ&Colliding \\
			\hline
			12715 (7275)  & 0 & 0 & 93 & 20 &0 & 1 & SEEB&  O-ring \\	
			\hline
		\end{tabular}
	\end{center}
	\caption{Parameters leading to categorization of $\delta$-spots.  Columns 2-5, the AH, AJ, Do$\delta$ and $\Delta$Rot values, are determined from the $\delta$-umbrae only, or the 'knot', as shown in Table 1. The $\Delta$Rot is for individual knots which is positive (negative) if it is counter-clockwise (clockwise), an $\ast$ indicates an average of several knots rotating in both directions, and $-$ indicates there was no measurable rotation. Column 6 indicates if it is a quadrupole (1, yes). Column 7 indicates if there are single or multiple flux emergence events, defined as emergence events separated by more than 48 hours that increase the flux by more than 30$\%$.  Column 8 is the first categorization indicating if the regions as formed with SEEB, SEEQ or MEEQ. Column 9 is the second categorization indicating which regions observational signatures are consistent with the formation mechanisms seen in Figure 1. 
	Categorization 1 shows that 58$\%$ are SEEQ, 26$\%$ SEEB and 16$\%$ MEEQ.  Categorization 2 shows that 42$\%$ have signatures consistent with the kink (or inverted kink) instability or $\Sigma$-event, 32$\%$ with multi-segment buoyancy, 16$\%$ with collisions with 11$\%$ unclassified but consistent with O-rings. If the quadrupole shows a higher rotation of the central umbrae in $\delta$-configuration ($\ge$90$\degree$), we classify it as inverted  kink configuration with (Quadrupole) as a secondary classification and vice versa if the rotation is less.}
\end{table}

\begin{figure}[b]
\begin{center}
\includegraphics[trim=1.1in 21.9in 1.7in 2.8in,clip,width=1.0\textwidth]{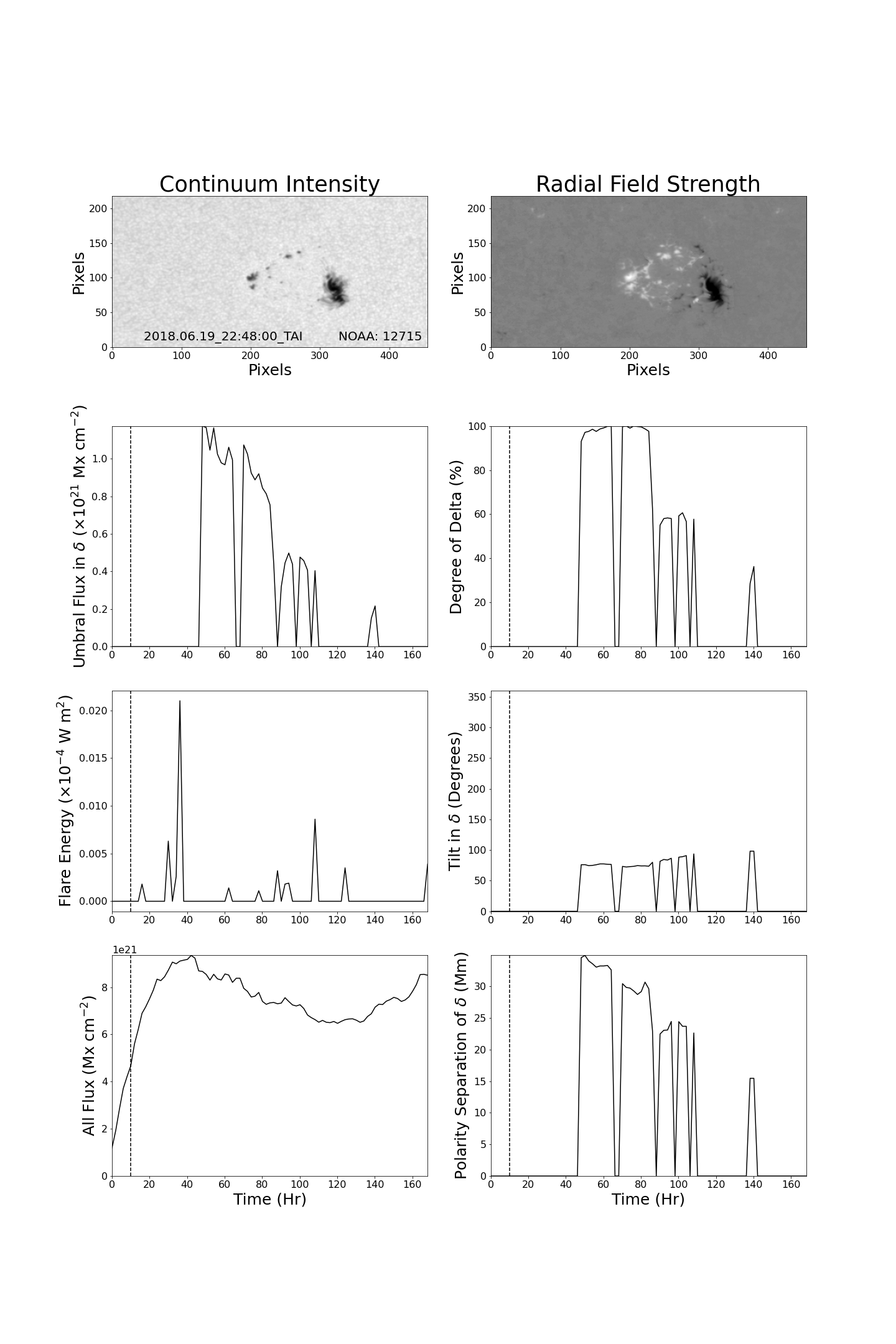}
\includegraphics[trim=1.1in 21.9in 1.7in 4.38in,clip,width=1.0\textwidth]{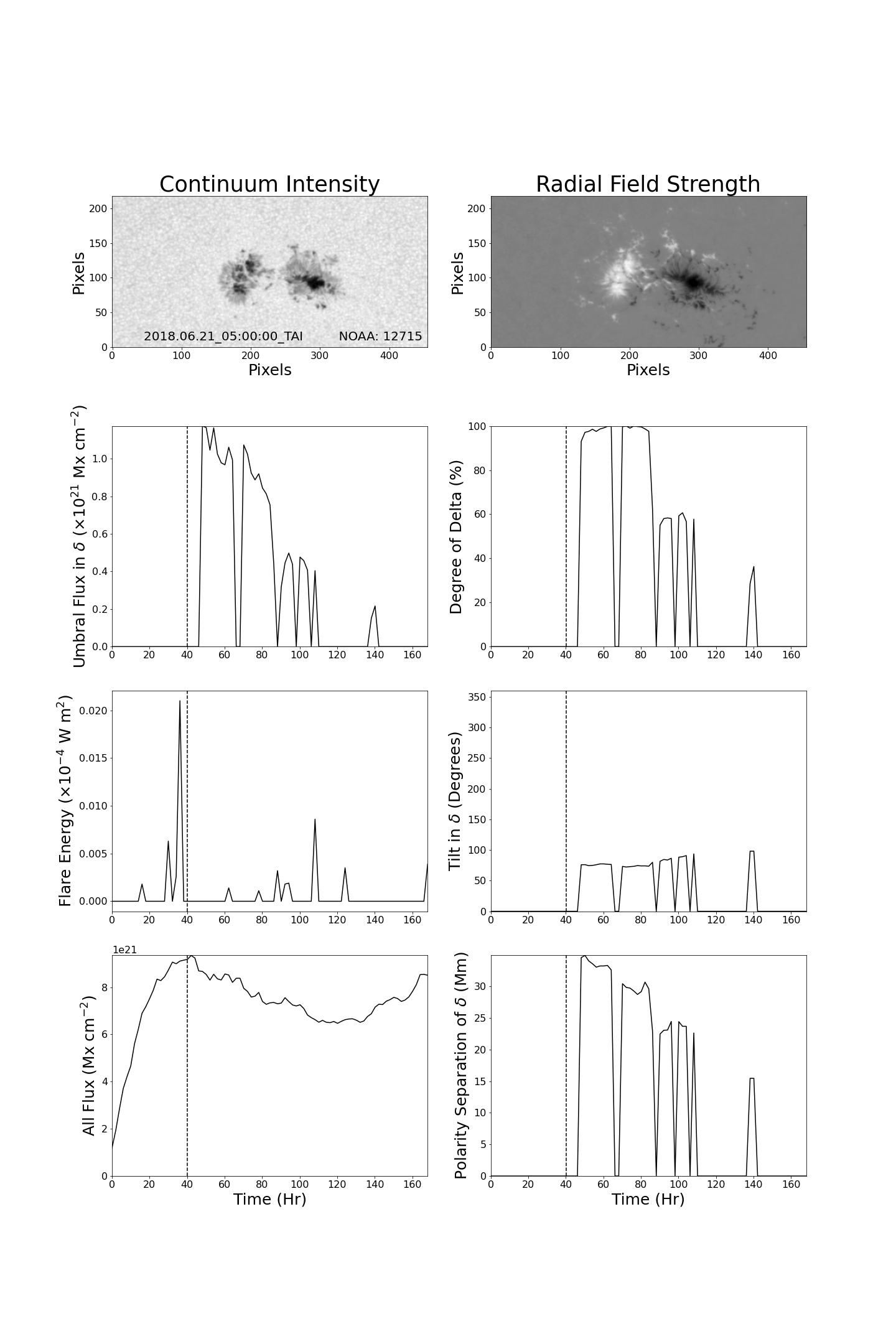}
\includegraphics[trim=1.1in 21.9in 1.7in 4.38in,clip,width=1.0\textwidth]{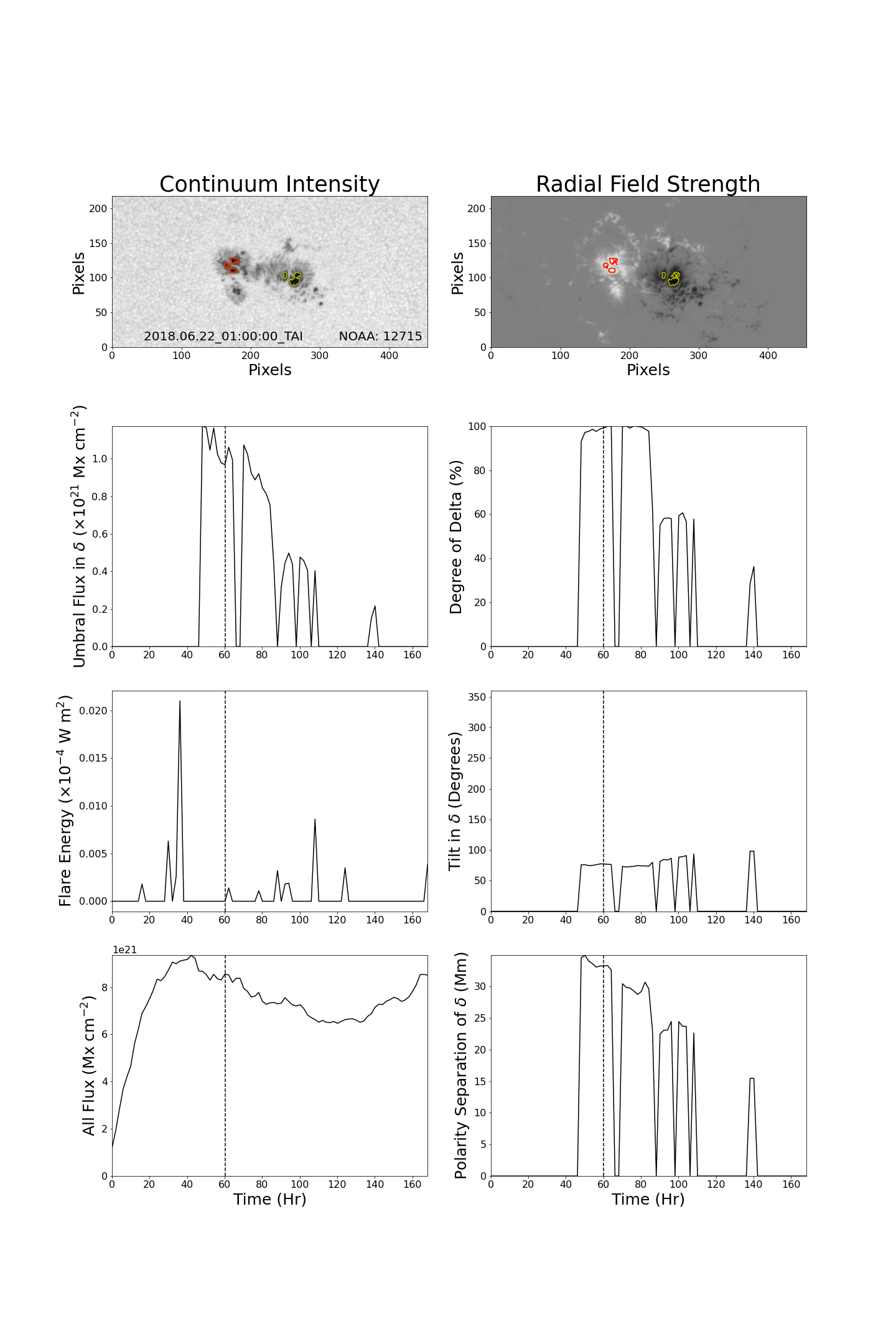}
\includegraphics[trim=1.1in 21.9in 1.7in 4.38in,clip,width=1.0\textwidth]{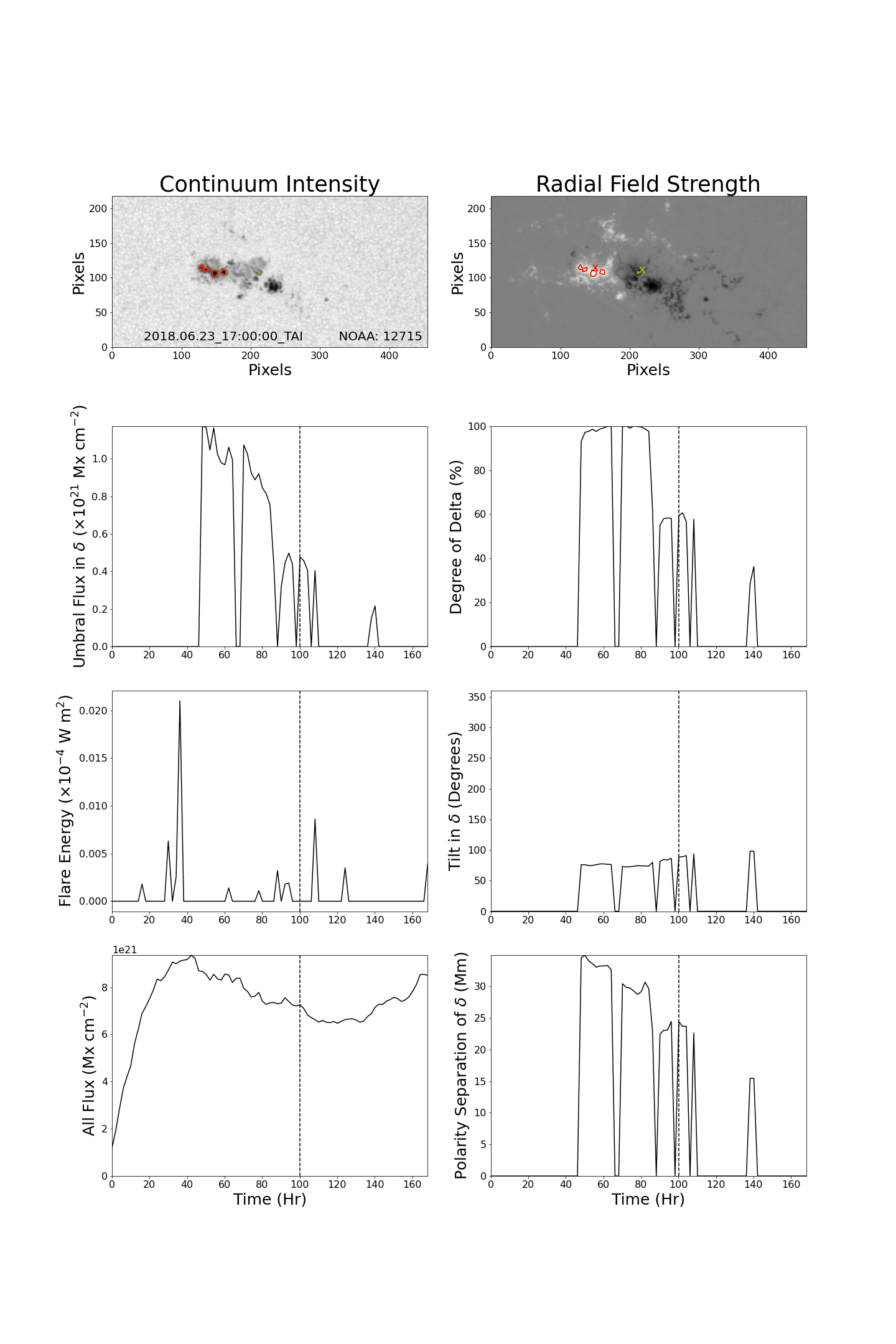}
\includegraphics[trim=1.1in 21.5in 1.7in 4.38in,clip,width=1.0\textwidth]{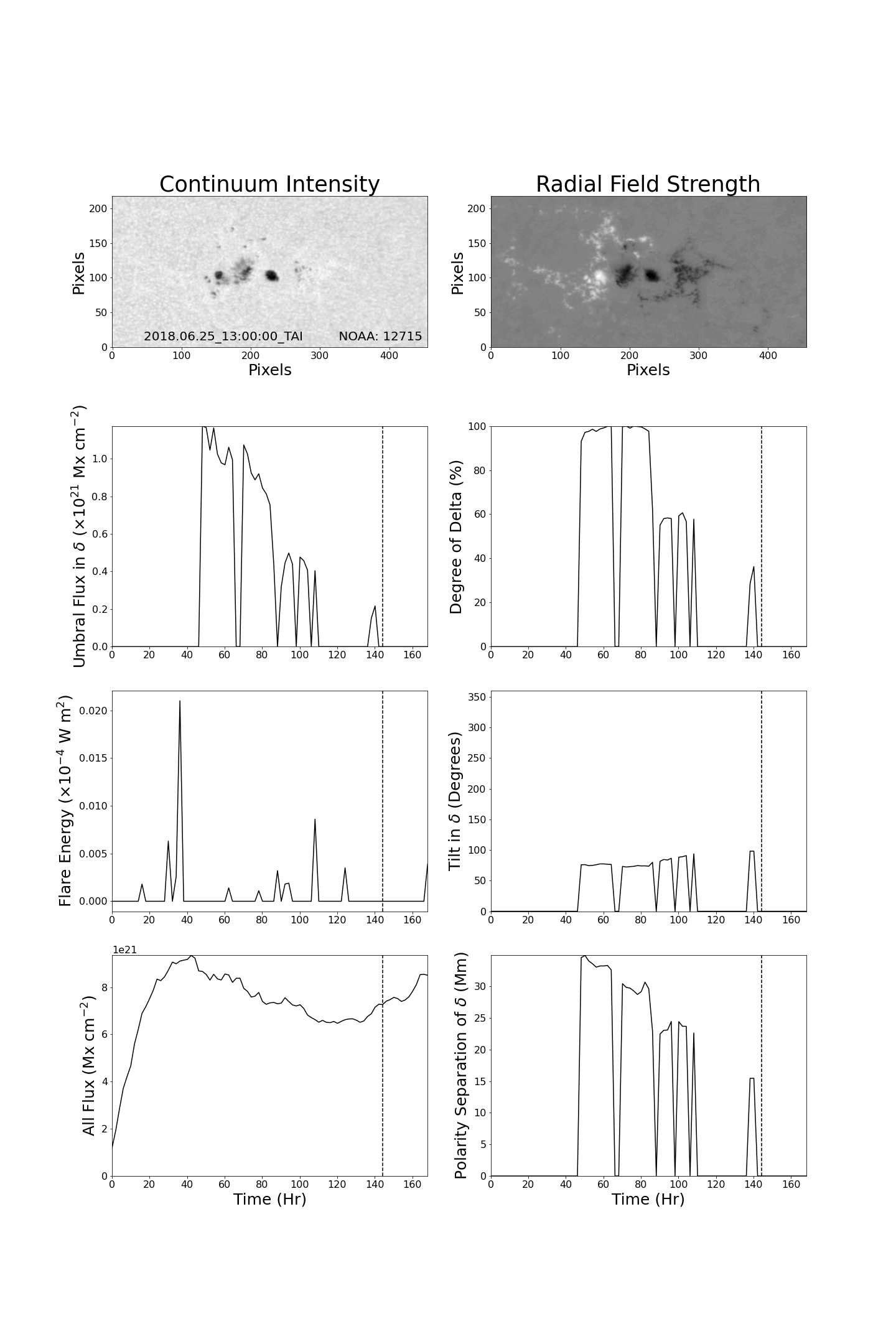}
\end{center}
\caption{NOAA 12715 snapshots are shown because its configuration and behavior is somewhat anomalous for a $\delta$-spot. The region is not AH or AJ. It is a single emergence event bipole (SEEB) with very little rotation whose opposite polarities emerge with some separation but remain close, converging somewhat over time.  We speculate that this could be the signature of flux emerging as an $\Omega$-loop initially, but the legs of the region have reconnected beneath in an O-ring and hence the polarities converge, or the region is confined by strong sub-surface flows.  Neither mechanism is considered to be a formation mechanism for $\delta$-spots. NOAA AR 11465 is another AR consistent with an O-ring.}
\label{Uloop}
\end{figure}

\begin{acknowledgments}
This work was supported by NASA HSR grant NNH18ZDA001N and NASA DRIVE Center COFFIES grant 80NSSC20K0602.
\end{acknowledgments}
\newpage
\newpage
\bibliography{delta}{}
\bibliographystyle{aasjournal}

\newpage
\appendix
\section{Delta Sunspots in Cycle 24 Observed by HMI}

This Appendix contains lists of all $\delta$-regions in Solar Cycle 24 as classified by NOAA observers.  The regions are found in Tables 5-14 organized by year with NOAA numbers and corresponding HARP number, date of observations, classification and maximum single flare energy. To put the $\delta$-regions in context of the progression of the solar cycle, we plot their location on a butterfly diagram created using HMI SHARP data, see Figure 12.  

\begin{figure}[hpt]
\includegraphics[trim=0.0in 0.10in 0.0in 0.1in,clip,width=0.95\textwidth]{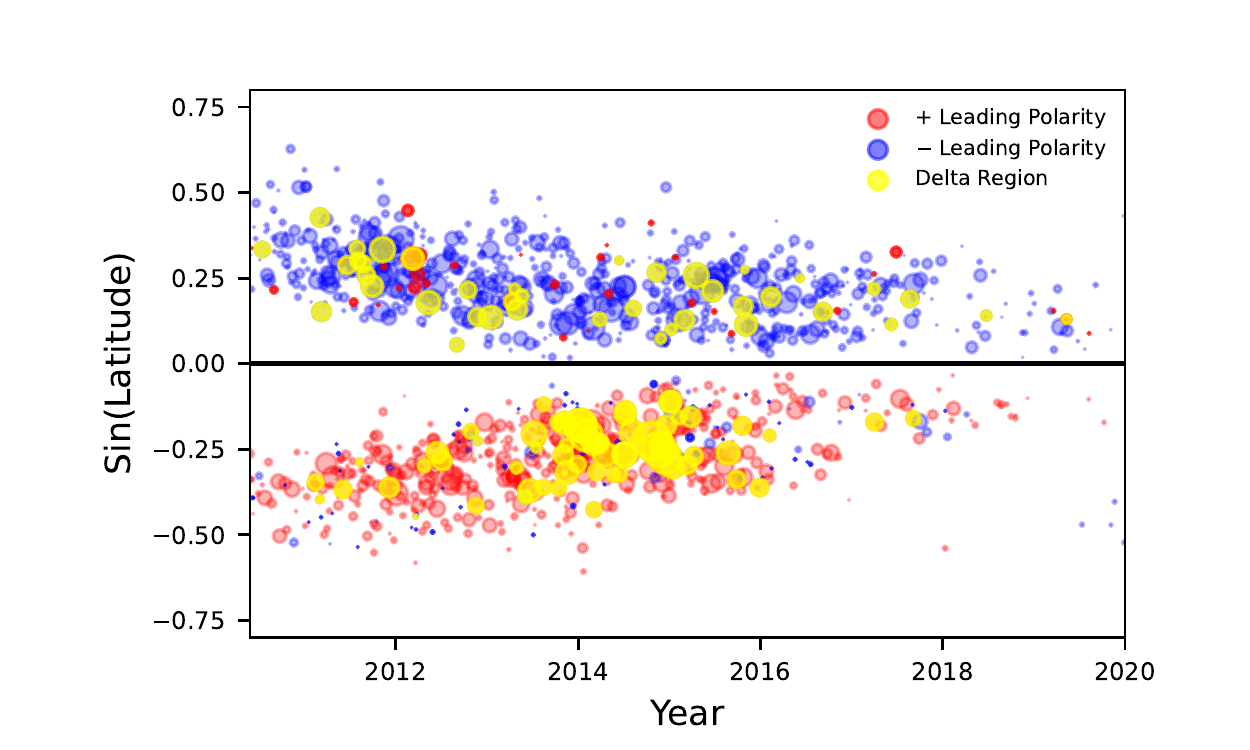}
\caption{The $\delta$-regions listed in Tables 5-14 are shown on a butterfly diagram where ARs are plotted as a function of time and sine latitude. $\delta$-regions are shown in yellow and are not colored for polarity or anti-Hale orientation. All non-$\delta$ regions observed during this cycle are also plotted, with the leading spot polarity is shown in red (blue) with total flux indicated by symbol marker size. Anti-Hale regions are obvious as the non-dominant color in each hemisphere.  The data used to generate this figure is the HMI SHARPS data summarized in the Solar Photospheric Ephemeral and Active Region (SPEAR) catalogue \citep{Norton:2021} that is an easy-to-read tabulated text file, SPEAR-CR.txt, available here: \textit{http://sun.stanford.edu/$\sim$norton/SPEAR/}. The catalogue currently contains information on nearly four thousand magnetic regions at their nearest central meridian crossing time for Carrington Rotations 2096 – 2239. For reference, the northern hemispheric sunspot number peaked in late 2011 and the southern hemispheric sunspot number peaked in early 2014, information available from multiple sources and found at \textit{http://users.telenet.be/j.janssens/SC24web}.  }
\label{bfly}
\end{figure}

\begin{longtable}{|c|c|l|l|l|p{4cm}|p{4.1cm}|}
\caption{Delta Sunspots of 2010} \label{tab:long} \\

\hline \multicolumn{1}{|c|}{\textbf{NOAA }} &
\multicolumn{1}{c|}{\textbf{HARP}} &
\multicolumn{1}{l|}{\textbf{Date}} & \multicolumn{1}{l|}{\textbf{Class}}  & \multicolumn{1}{l|}{\textbf{Flare}} \\ \hline 
11045  & - & 20100209 - 10& $\beta\gamma\delta$ & C3 \\
11087 & 86 & 20100713 & $\beta\delta $ & C2\\
\hline
\end{longtable}
\begin{longtable}{|c|c|l|l|l|p{4cm}|p{4.1cm}|}
\caption{Delta Sunspots of 2011} \label{tab:long} \\

\hline \multicolumn{1}{|c|}{\textbf{NOAA }} &
 \multicolumn{1}{c|}{\textbf{HARP}} &
 \multicolumn{1}{l|}{\textbf{Date}} & \multicolumn{1}{l|}{\textbf{Class}}  & \multicolumn{1}{l|}{\textbf{Flare}} \\ \hline 
11158 &377 & 20110216 - 19 & $\beta\gamma\delta$ & X2\\
11161 &384 & 20110220 & $\beta\gamma\delta$ & C4\\
11164 &393 & 20110302 - 09 & $\beta\gamma\delta$ & C8\\
11165 &394 & 20110308 - 09 & $\beta\gamma\delta$ & C5\\
11166 &401 & 20110308 - 13 & $\beta\gamma\delta$ & X1\\
\hline
11224 &622 & 20110530 & $\beta\delta$ & C3\\
11226 &637 & 20110529 - 30 & $\beta\delta$ & C4\\
11236 &667 & 20110616 - 17 & $\beta\delta$ & C7\\
11260 &746 & 20110729 & $\beta\gamma\delta$ & M1\\
11261 &750 & 20110731 - 0805 & $\beta\gamma\delta$ & M9\\
\hline
*11262 &744 & 20110728 & $\beta\gamma\delta$ & -\\
11263 &753 & 20110802 - 10 & $\beta\gamma\delta$ & M1\\
11267 &764 & 20110807 & $\beta\gamma\delta$ & C1\\
11271 &794 & 20110818 - 26 & $\beta\gamma\delta$ & C2\\
*11282 &814 & 20110904 & $\beta\delta$ & -\\
\hline
11283 &833 & 20110907 - 09 & $\beta\gamma\delta$ & X2\\
11302 &892 & 20110925 - 1003 & $\beta\gamma\delta$ & M4\\
11339 &1028 & 20111104 - 08 & $\beta\gamma\delta$ & M1\\
11363 &1124 & 20111205 & $\beta\gamma\delta$ & C6\\
11374 &1168 & 20111213 & $\beta\delta$ & C1\\
\hline
\end{longtable}
\begin{longtable}{|c|c|l|l|l|p{4cm}|p{4.1cm}|}
\caption{Delta Sunspots of 2012} \label{tab:long} \\

\hline \multicolumn{1}{|c|}{\textbf{NOAA }} &
\multicolumn{1}{c|}{\textbf{HARP}} &
\multicolumn{1}{l|}{\textbf{Date}} & \multicolumn{1}{l|}{\textbf{Class}}  & \multicolumn{1}{l|}{\textbf{Flare}} \\ \hline
11429 &1449  & 20120305 - 12 & $\beta\gamma\delta$ & X5\\
11440 &1484  & 20120322 & $\beta\gamma\delta$ & C2\\
11465 &1596  & 20120425 - 28 & $\beta\gamma\delta$ & C2\\
11476 &1638  & 20120509 - 14 & $\beta\gamma\delta$ & M5\\
11504 &1750  & 20120614 - 15 & $\beta\gamma\delta$ & M1\\
\hline
11515 &1807 & 20120704 - 07 & $\beta\gamma\delta$ & M5\\
11520 &1834  & 20120710 - 16 & $\beta\gamma\delta$ & X1\\
11560 &1993  & 20120904 - 06 & $\beta\gamma\delta$ & C5\\
11589 &2109 & 20121014 & $\beta\gamma\delta$ & C3\\
11598 &2137  & 20121024 - 28 & $\beta\delta$ & C4\\
\hline
11613 &2191  & 20121114 & $\beta\gamma\delta$ & M6\\
11618 &2220 & 20121122 - 28 & $\beta\gamma\delta$ & M3\\
11620 &2227  & 20121128 - 30 & $\beta\gamma\delta$ & M1\\

\hline
\end{longtable}
\begin{longtable}{|c|c|l|l|l|p{4cm}|p{4.1cm}|}
\caption{Delta Sunspots of 2013} \label{tab:long} \\

\hline \multicolumn{1}{|c|}{\textbf{NOAA }} &
\multicolumn{1}{c|}{\textbf{HARP}} &
\multicolumn{1}{l|}{\textbf{Date}} & \multicolumn{1}{l|}{\textbf{Class}}  & \multicolumn{1}{l|}{\textbf{Flare}} \\ \hline
11640 &2337 & 20130104 - 06 & $\beta\gamma\delta$ & C1\\
11654 &2372 & 20130116 - 17 & $\beta\gamma\delta$ & C5\\
11678 &2469 & 20130220 - 22 & $\beta\gamma\delta$ & C8\\
11719 &2635 & 20130412 & $\beta\gamma\delta$ & M6\\
11726 &2673 & 20130422 - 26 & $\beta\gamma\delta$ & M1\\
\hline
11730 &2691 & 20130430 - 0502 & $\beta\gamma\delta$ & C9\\
11731 &2693 & 20130501 - 04 & $\beta\gamma\delta$ & M1\\
11748 &2748 & 20130515 - 21 & $\beta\gamma\delta$ & M3\\
11762 &2790 & 20130604 - 06 & $\beta\gamma\delta$ & C9\\
11775 &2852 & 20130618 - 22 & $\beta\delta$ & C2\\
\hline
11785 &2920 & 20130705 - 11 & $\beta\gamma\delta$ & C9\\
11787 &2920 & 20130709 & $\beta\gamma\delta$ & C4\\
11791 &2952 & 20130716 - 17 & $\beta\gamma\delta$ & C3\\
11817 &3048 & 20130813 - 18 & $\beta\gamma\delta$ & M1\\
11818 &3056 & 20130817 - 19 & $\beta\gamma\delta$ & M3\\
\hline
11861 &3258, 3263 & 20131012 & $\beta\gamma\delta$ & C7\\
11865 &3263 & 20131010 - 17 & $\beta\gamma\delta$ & M1\\
11875 &3291 & 20131022 - 30 & $\beta\gamma\delta$ & M4\\
11877 &3296 & 20131024 - 28 & $\beta\gamma\delta$ & M9\\
11882 &3311 & 20131026 - 29 & $\beta\gamma\delta$ & M4\\
\hline
11884 &3321 & 20131028 - 1103 & $\beta\gamma\delta$ & M6\\
11890 &3341 & 20131105 - 13 & $\beta\gamma\delta$ & X1\\
11891 &3344 & 20131108 - 09 & $\beta\delta$ & M2\\
11893 &3364 & 20131118 - 20 & $\beta\delta$ & C4\\
11897 &3366 & 20131116 & $\beta\gamma\delta$ & C8\\
\hline
11934 &3520 & 20131229 - 140101 & $\beta\gamma\delta$ & C2\\
11936 &3535 & 20131231 - 140102 & $\beta\gamma\delta$ & M6\\

\hline
\end{longtable}
\begin{longtable}{|c|l|l|l|l|p{4cm}|p{4.1cm}|}
\caption{Delta Sunspots of 2014} \label{tab:long} \\
\hline \multicolumn{1}{|c|}{\textbf{NOAA }} & 
\multicolumn{1}{l|}{\textbf{HARP}} &
\multicolumn{1}{l|}{\textbf{Date}} &
\multicolumn{1}{l|}{\textbf{Class}}  & \multicolumn{1}{l|}{\textbf{Flare}} \\ \hline
11944 &3563 & 20140105 - 12 & $\beta\gamma\delta$ & X1\\
11967 &3686 & 20140130 - 0210 & $\beta\gamma\delta$ & M5\\
11974 &3721 & 20140212 - 18 & $\beta\gamma\delta$ & M3\\
11990 &3793 & 20140227 - 0303 & $\beta\delta$ & C2\\
11991 &3804 & 20140305 & $\beta\gamma\delta$ & M1\\
\hline
12002 &3836 & 20140310 - 14 & $\beta\gamma\delta$ & C9\\
12010 &3856 & 20140324 - 25 & $\beta\gamma\delta$ & M1\\
12015 &3856 & 20140324 - 26 & $\beta\delta$ & C3\\
12017 &3894 & 20140329 - 30 & $\beta\gamma\delta$ & X1\\
12021 &3912 & 20140405 & $\beta\gamma\delta$ & C6\\
\hline
12035 &4000 & 20140419 & $\beta\gamma\delta$ & M1\\
12051 &4071 & 20140503 - 07 & $\beta\gamma\delta$ & C5\\
12056 &4097 & 20140509 & $\beta\gamma\delta$ & C8\\
12065 &4138 & 20140526 & $\beta\gamma\delta$ & C1\\
12080 &4197 & 20140607 - 15 & $\beta\gamma\delta$ & M1\\
\hline
12085 &4197 & 20140610 - 13 & $\beta\gamma\delta$ & C9\\
12087 &4225 & 20140611 - 14 & $\beta\gamma\delta$ & M2\\
12089 &4231 & 20140616 - 17 & $\beta\gamma\delta$ & M1\\
12104 &4296 & 20140701 - 03 & $\beta\gamma\delta$ & C1\\
12107 &4296 & 20140702 & $\beta\delta$ & C2\\
\hline
12108 &4315 & 20140707 - 10 & $\beta\gamma\delta$ & C4\\
12109 &4321 & 20140707 - 10 & $\beta\gamma\delta$ & C4\\
12127 &4396 & 20140729 - 30 & $\beta\delta$ & M1\\
12130 &4396 & 20140731 - 0804 & $\beta\gamma\delta$ & M2\\
12132 &4396 & 20140802 - 03 & $\beta\gamma\delta$ & C2\\
\hline
12134 &4424 & 20140806 - 07 & $\beta\gamma\delta$ & -\\
12146 &4466 & 20140826 - 29 & $\beta\gamma\delta$ & M3\\
12149 &4477 & 20140825 & $\beta\gamma\delta$ & C4\\
12157 &4530 & 20140906 - 12 & $\beta\gamma\delta$ & C9\\
12158 &4536 & 20140907 - 10 & $\beta\gamma\delta$ & X1\\
\hline
12172 &4580 & 20140922 & $\beta\delta$ & M2\\
12175 &4591 & 20140927 - 30 & $\beta\gamma\delta$ & C5\\
12192 &4698 & 20141020 - 30 & $\beta\gamma\delta$ & X3\\
12205 &4781 & 20141106 - 12 & $\beta\gamma\delta$ & X1\\
12209 &4817 & 20141116 - 26 & $\beta\gamma\delta$ & C8\\
\hline
12216 &4851 & 20141122 - 24 & $\beta\gamma\delta$ & C2\\
12219 &4868 & 20141130 & $\beta\gamma\delta$ & C6\\
12241 &4941 & 20141218 - 22 & $\beta\gamma\delta$ & M6\\
12242 &4920 & 20141216 - 22 & $\beta\gamma\delta$ & X1\\

\hline
\end{longtable}
\begin{longtable}{|c|c|l|l|l|p{4cm}|p{4.1cm}|}
\caption{Delta Sunspots of 2015} \label{tab:long} \\
\hline \multicolumn{1}{|c|}{\textbf{NOAA }} &
\multicolumn{1}{c|}{\textbf{HARP}} &
\multicolumn{1}{l|}{\textbf{Date}} & \multicolumn{1}{l|}{\textbf{Class}}  & \multicolumn{1}{l|}{\textbf{Flare}} \\ \hline
12253 &5011 & 20150103 - 06 & $\beta\gamma\delta$ & M1\\
12255 &5022 & 20150112 & $\beta\gamma\delta$ & C1\\
12257 &5026 & 20150110 - 14 & $\beta\gamma\delta$ & C9\\
12259 &5039 & 20150115 - 16 & $\beta\gamma\delta$ & C2\\
12280 &5144 & 20150209 - 11 & $\beta\gamma\delta$ & C8\\
\hline
12293 &5249 & 20150228 & $\beta\delta$ & C4\\
12297 &5298 & 20150309 - 19 & $\beta\gamma\delta$ & X2\\
12305 &5354 & 20150325 - 26 & $\beta\gamma\delta$ & C8\\
12320 &5415 & 20150408 - 11 & $\beta\delta$ & M1\\
12321 &5447 & 20150413 - 14 & $\beta\gamma\delta$ & C7\\
\hline
12371 &5692 & 20150619 - 24 & $\beta\gamma\delta$ & M7\\
12403 &5885 & 20150823 - 30 & $\beta\gamma\delta$ & M5\\
12422 &5983 & 20150927 - 1003 & $\beta\gamma\delta$ & M7\\
12434 &6015 & 20151018 & $\beta\gamma\delta$ & C4\\
12436 &6027 & 20151021 - 22 & $\beta\delta$ & C7\\
\hline
12443 &6063 & 20151031 - 1110 & $\beta\gamma\delta$ & M3\\
12445 &6052 & 20151104 & $\beta\delta$ & C2\\
12473 &6206 & 20151223 - 29 & $\beta\gamma\delta$ & M1\\

\hline
\end{longtable}
\begin{longtable}{|c|c|l|l|l|p{4cm}|p{4.1cm}|}
\caption{Delta Sunspots of 2016} \label{tab:long} \\
\hline \multicolumn{1}{|c|}{\textbf{NOAA }} &
\multicolumn{1}{l|}{\textbf{HARP}} &
\multicolumn{1}{l|}{\textbf{Date}} & \multicolumn{1}{l|}{\textbf{Class}}  & \multicolumn{1}{l|}{\textbf{Flare}} \\ \hline
12494 &6320 & 20160205 - 08 & $\beta\gamma\delta$ & C5\\
12497 &6327 & 20160212 - 18 & $\beta\gamma\delta$ & M1\\
12552 &6599 & 20160610 - 11 & $\beta\delta$ & C6\\
12567 &6670 & 20160716 & $\beta\gamma\delta$ & C2\\
12585 &6731 & 20160909 - 11 & $\beta\delta$ & B3\\

\hline
\end{longtable}
\begin{longtable}{|c|c|l|l|l|p{4cm}|p{4.1cm}|}
\caption{Delta Sunspots of 2017} \label{tab:long} \\
\hline \multicolumn{1}{|c|}{\textbf{NOAA }} &
\multicolumn{1}{c|}{\textbf{HARP}} &\multicolumn{1}{l|}{\textbf{Date}} & \multicolumn{1}{l|}{\textbf{Class}}  & \multicolumn{1}{l|}{\textbf{Flare}} \\ \hline
12645 &6975 & 20170403 - 06 & $\beta\gamma\delta$ & C5\\
12644 &6972 & 20170404 & $\beta\gamma\delta$ & C3\\
12661 &7034 & 20170606 & $\beta\delta$ & B5\\
12671 &7107 & 20170817 & $\beta\gamma\delta$ & B6\\
12673 &7115 & 20170905 - 10 & $\beta\gamma\delta$ & X9\\

\hline
\end{longtable}
\begin{longtable}{|c|c|l|l|l|p{4cm}|p{4.1cm}|}
\caption{Delta Sunspots of 2018} \label{tab:long} \\
\hline \multicolumn{1}{|c|}{\textbf{NOAA }} &
\multicolumn{1}{c|}{\textbf{HARP}} &
\multicolumn{1}{l|}{\textbf{Date}} & \multicolumn{1}{l|}{\textbf{Class}}  & \multicolumn{1}{l|}{\textbf{Flare}} \\ \hline
12715 &7275 & 20180624 & $\beta\delta$ & B8\\

\hline
\end{longtable}
\begin{longtable}{|c|c|l|l|l|p{4cm}|p{4.1cm}|}
\caption{Delta Sunspots of 2019} \label{tab:long} \\
\hline \multicolumn{1}{|c|}{\textbf{NOAA }} & 
\multicolumn{1}{l|}{\textbf{HARP}} &
\multicolumn{1}{l|}{\textbf{Date}} & \multicolumn{1}{l|}{\textbf{Class}}  & \multicolumn{1}{l|}{\textbf{Flare}} \\ \hline
12736 &7350 & 20190322 & $\beta\gamma\delta$ & C4\\
12740 &7357 & 20190506 - 07 & $\beta\delta$ & M1\\
\hline
\end{longtable}

\end{document}